\documentclass[twocolumn]{aastex63}
\usepackage{longtable,morefloats}
\usepackage{apjfonts}
\usepackage{mathrsfs}
\usepackage{booktabs}
\usepackage{hyperref}
\usepackage{amsmath}
\usepackage{natbib}
\usepackage{float}
\usepackage{bm}

\newcommand\iona[2]{#1$\;${\scshape{#2}}}
\newcommand{\xray}{{\hbox{X-ray}}}

\begin{document}

\title{Rest-Frame Optical Spectroscopy of Ten \MakeLowercase{$z \sim 2$} Weak Emission-Line Quasars
}
\author{Y.~Chen}
\affiliation{School of Astronomy and Space Science, Nanjing University, Nanjing, Jiangsu 210093, China}
\affiliation{Key Laboratory of Modern Astronomy and Astrophysics (Nanjing University), Ministry of Education, Nanjing 210093, China}

\author{B.~Luo}
\affiliation{School of Astronomy and Space Science, Nanjing University, Nanjing, Jiangsu 210093, China}
\affiliation{Key Laboratory of Modern Astronomy and Astrophysics (Nanjing University), Ministry of Education, Nanjing 210093, China}

\author{W.~N.~Brandt}
\affiliation{Department of Astronomy \& Astrophysics, 525 Davey Lab,
The Pennsylvania State University, University Park, PA 16802, USA}
\affiliation{Institute for Gravitation and the Cosmos, The Pennsylvania State University,
University Park, PA 16802, USA}
\affiliation{Department of Physics, 104 Davey Lab, The Pennsylvania State University, University Park, PA 16802, USA}

\author{Wenwen~Zuo}
\affiliation{Shanghai Astronomical Observatory, Chinese Academy of Sciences, 80 Nandan Road, Shanghai 200030, China}

\author{Cooper~Dix}
\affiliation{Department of Physics, University of North Texas, Denton, TX, 76203, USA}
\affiliation{Department of Astronomy, The University of Texas at Austin, Austin, TX 78712, USA}

\author{Trung~Ha}
\affiliation{Department of Physics, University of North Texas, Denton, TX, 76203, USA}

\author{Brandon~Matthews}
\affiliation{Department of Physics, University of North Texas, Denton, TX, 76203, USA}

\author{Jeremiah~D.~Paul}
\affiliation{Department of Physics, University of Nevada, Reno, NV 89557, USA}

\author{Richard~M.~Plotkin}
\affiliation{Department of Physics, University of Nevada, Reno, NV 89557, USA}
\affiliation{Nevada Center for Astrophysics, University of Nevada, Las Vegas, NV 89154, USA}

\author{Ohad~Shemmer}
\affiliation{Department of Physics, University of North Texas, Denton, TX, 76203, USA}



\begin{abstract}
We present near-infrared spectroscopy of ten weak \hbox{emission-line} quasars (WLQs)
at redshifts of $z\sim2$, obtained with the Palomar 200-inch Hale Telescope.
WLQs are an exceptional population of type 1 quasars that exhibit weak or 
no broad emission lines in the ultraviolet (e.g., the \iona{C}{iv}~$\lambda 1549$ line),
and they display remarkable \xray\ properties.
We derive H$\beta$-based \hbox{single-epoch} virial black-hole masses (median value 
$\rm 1.7 \times 10^{9}~M_{\odot}$) and Eddington ratios (median value $0.5)$ for our sources.
We confirm the previous finding that WLQ H$\beta$ lines, 
as a major \hbox{low-ionization} line, are not significantly weak compared to typical quasars.
The most prominent feature of the WLQ optical spectra is 
the universally weak/absent [\iona{O}{iii}]~$\lambda 5007$ emission.
They also display stronger optical \iona{Fe}{ii} emission than typical quasars.
Our results favor the \hbox{super-Eddington} accretion scenario for WLQs, 
where the weak lines are a result of a soft ionizing continuum; 
the geometrically thick inner accretion disk
and/or its associated outflow is responsible for obscuring the nuclear 
high-energy radiation and producing the soft ionizing continuum.
We also report candidate extreme [\iona{O}{iii}] outflows 
(blueshifts of $\approx~500$ and $\rm 4900~km~s^{-1}$) in one object.
\end{abstract}


\section{Introduction}
\label{sec:intro}
The spectra of type 1 quasars are characterized by strong and broad emission lines in the optical and ultraviolet (UV).
These lines are considered to be produced by high-velocity photoionized gas in the broad \hbox{emission-line} region (BELR) that is within the gravitational potential well of the central supermassive black hole (SMBH).
The size of the BELR, derived either from reverberation-mapping campaigns or an empirical luminosity--radius relation \citep[e.g.,][]{Maoz1991,Peterson1991,Kaspi2000,bentz2009,Du2014}, combined with the virial velocity of the BELR obtained from \hbox{emission-line} profiles, provide an effective approach to estimate the masses of SMBHs in distant quasars \citep[e.g.,][and references therein]{Shenliu2012}. 

A small population (a few hundred) of type 1 quasars have been discovered to exhibit exceptionally weak or no broad emission lines in the UV \citep[e.g., Ly$\alpha$~$+$~\iona{N}{v}~$\lambda$1240 complex, \iona{C}{iv}~$\lambda 1549$, \iona{Mg}{ii}~$\lambda 2799$;][]{Fan1999,Diamond-Stanic2009,Plotkin2010,Shemmer2010}.
These weak \hbox{emission-line} quasars (WLQs) have generally been found at high redshifts ($z\gtrsim 1.5$), where optical spectroscopy covers the \hbox{rest-frame} UV spectrum.
There is no uniform \hbox{emission-line} strength definition for WLQs. 
Typically they are selected to have \hbox{rest-frame} equivalent widths (REWs) $\approx3$--$4\sigma$ below the mean of the REW distribution for the considered UV line. 
For example, REW(Ly$\alpha$~$+$~\iona{N}{v})$<15.4~\textup{\AA}$ and REW(Ly$\alpha$~$+$~\iona{N}{v})$<10~\textup{\AA}$ were adopted in \citet{Diamond-Stanic2009} and \citet{Shemmer2009}, respectively, and the lower $3\sigma$ limit of the log-normal \iona{C}{iv} REW distribution is $\approx10~\textup{\AA}$ \citep[e.g.,][]{Diamond-Stanic2009,Wu2012}.
It also appears difficult to unify WLQ populations identified with different lines \citep[e.g.,][]{Wu2012,Paul2022}. 
The fraction of WLQs likely increases with redshift. 
Based on a sample of $\sim3000$ Sloan Digital Sky Survey (SDSS; \citealt{York2000}) quasars in the redshift range of $\sim3$--$5$, \citet{Diamond-Stanic2009} found that the WLQ fraction is $1.3\%$ (36/2737) below $z\sim4.2$, and it is $6.2\%$ (20/321) at $z\gtrsim4.2$.
Of 117 Pan-STARRS1 quasars at $5.6\lesssim z\lesssim 6.7$, \citet{Banados2016} identified 16 (13.7\%) WLQs.
Several of the ultraluminous quasars at $z > 4.5$ are also WLQs \citep[e.g.,][]{Wuxb2015,Wolf2020}.
Besides the exceptional \hbox{emission-line} properties, WLQs have typical quasar continuum spectral energy distributions (SEDs) from the infrared (IR) to UV \citep[e.g.,][]{Diamond-Stanic2009,Shemmer2010,Lane2011,Wu2011,Wu2012,Luo2015,Plotkin2015}.  

The physical mechanism responsible for the remarkable \hbox{emission-line} properties of WLQs is not clear. 
The weak line emission is not caused by contamination from jet emission, line obscuration, gravitational lensing effects, or broad absorption line (BAL) effects. 
One possible explanation is that the WLQ BELR is gas deficient due to low gas content and/or a small covering factor \citep[the anemic BELR scenario; e.g.,][]{Shemmer2010}. 
Another possible explanation is that the weak line emission is caused by a “soft” nuclear ionizing continuum that cannot ionize the BELR to the typical ionization state. 
In the former scenario, one might expect that all broad emission lines of WLQs would be weak. 
However, simultaneous \hbox{rest-frame} optical and UV spectroscopy of a small sample of six WLQs with VLT/X-Shooter revealed that only the UV \iona{C}{iv} lines are substantially weaker (at $\rm > 3\sigma$ levels) than those of typical quasars, while the optical lines such as H$\beta$~$\lambda 4861$ are not significantly weak compared to typical quasars \citep{Plotkin2015}. 
This result favors the soft ionizing continuum scenario where only the \hbox{high-ionization} BELR is strongly affected, leading to the exceptionally weak \iona{C}{iv} emission. 
Rest-frame optical spectroscopy of a large sample of WLQs is needed to investigate systematically the strength of their $\rm H\beta$ lines and assess whether the soft ionizing continuum scenario is the most probable explanation for WLQs.

Despite the typical IR--UV continuum SEDs of WLQs, they show remarkable \xray\ properties. 
Systematic investigations of the \xray\ emission from WLQs have revealed that a significant fraction ($\approx 30~\textrm{--}~50\%$) of them are also distinctly \xray\ weak compared to typical quasars \citep{Wu2011,Wu2012,Luo2015,Ni2018,Ni2022,Pu2020}.
The level of \xray\ weakness is often assessed by the deviation from the empirical $\alpha_{\rm OX}$--$L_{\rm 2500~{\textup{\AA}}}$ relation \citep[e.g.,][]{steffen2006},
\footnote{$\alpha_{\rm OX}$
is defined as
$\alpha_{\rm OX}=-0.3838\log(f_{2500~{\textup{\AA}}}/f_{2~{\rm keV}}$),
with $f_{2500~{\textup{\AA}}}$ and $f_{2~{\rm keV}}$
being the \hbox{rest-frame}
$2500~{\textup{\AA}}$ and $2~{\rm keV}$ flux densities, respectively. 
$L_{\rm 2500~{\textup{\AA}}}$ is the $\rm 2500~{\textup{\AA}}$ 
monochromatic luminosity.} 
which describes a negative correlation between the relative \xray\ emission strength and the optical/UV luminosity. 
The average \xray\ weakness factor for the \xray\ weak WLQs is $\approx13$, and their stacked effective \hbox{power-law} photon index ($\Gamma_{\rm eff}$) is flatter ($1.1^{+0.2}_{-0.1}$) than that of the \xray\ normal WLQs ($1.8\pm0.1$), suggestive of \xray\ obscuration \citep[e.g.,][]{Luo2015,Ni2022}.
A few WLQs have been observed to vary between \xray\ weak and \xray\ nominal-strength states \citep[e.g.,][]{Miniutti2012,Ni2020,Ni2022,Liu2022}, and the sometimes rapid variability even suggested an origin of a few tens of gravitational radii \citep{Liu2022}.

Based on the \xray\ results and the soft ionizing continuum scenario, it has been proposed that WLQs probably have \hbox{super-Eddington} accretion rates, and a geometrically Thick inner accretion Disk and/or its associated Outflow (TDO) is responsible for obscuring the nuclear high-energy radiation and producing a soft ionizing continuum (e.g., \citealt{Luo2015,Ni2018,Ni2022}; see the schematic diagram in Figure~1 of \citealt{Ni2018}). 
In this scenario, the quasar emits nominal levels of IR-to-X-ray radiation (i.e., typical quasar continuum SED) from the torus, accretion disk, and corona.
However, the TDO shields the equatorial high-ionization BELR, and thus the ionizing continuum received by the BELR is soft, causing the weak emission lines such as the \iona{C}{iv} line.
Moreover, if the inclination angle of the quasar is large (i.e., a more edge-on view),
and our line of sight intercepts the TDO,
we would observe an \xray\ obscured (\xray\ weak) quasar.
The obscuring TDO is probably clumpy,
resulting in the (sometimes rapid) \xray\ variability observed in a few WLQs.
The IR--UV continuum SED is not greatly affected in this scenario.
This TDO model can explain, in a simple and unified manner, the weak emission lines, remarkable \xray\ properties, and other multiwavelength properties of WLQs.
However, it is generally difficult to obtain reliable Eddington-ratio estimates for quasars.
The situation becomes worse for high-redshift quasars with only UV spectra, as the \iona{C}{iv}-based virial SMBH masses are likely biased or even incorrect \citep[e.g.,][]{Baskin2005,Denney2012,Kratzer2015,Coatman2017,Dix2023}.
Some of the ``Eigenvector 1'' parameters have been suggested to be correlated with the accretion rate \citep[e.g.,][]{Boroson1992,Sulentic2000,Shen2014}, but measurements of these also require \hbox{rest-frame} optical spectroscopy.

In this study, we present near-IR (NIR) spectra of a sample of ten WLQs observed with the Palomar Hale 200-inch telescope (P200), expanding the currently limited samples of WLQs with \hbox{rest-frame} optical spectra \citep[e.g.,][]{Shemmer2010,Wu2011,Plotkin2015,Shemmer2015,Ha2023}.
We examine the optical \hbox{emission-line} properties of these WLQs and evaluate the TDO scenario.
The paper is organized as follows.
We describe the sample selection, P200 data reduction, spectral analysis, and high-$z$ comparison samples in Section~\ref{sec:data}.
In Section~\ref{sec:result}, we present the \hbox{emission-line} properties and relevant comparisons; we also provide SMBH mass and Eddington-ratio estimates.
We discuss our results in the context of the TDO scenario in Section~\ref{sec:dis}, 
and we report candidate extreme [\iona{O}{iii}]~$\rm \lambda 5007$ outflows in one of the sources.
We summarize our main results in Section~\ref{sec:sum}.
In the Appendix, we present the spectrum of a low-$z$ WLQ candidate that had an incorrect redshift in the SDSS quasar catalog.
Throughout this paper, we use a cosmology with $H_0=67.4$~km~s$^{-1}$~Mpc$^{-1}$,
$\Omega_{\rm M}=0.315$, $\Omega_{\Lambda}=0.685$ \citep{Planck2020}.
Uncertainties are quoted at a $1\sigma$ confidence level, and upper limits are at a $3\sigma$ confidence level.

\section{Sample Selection and Near-Infrared Spectroscopy} 
\label{sec:data}


\subsection{Sample Selection and P200/TSpec Observations}
\label{subsec:sample}
We first selected our WLQ targets from the SDSS Data Release 7 (DR7; \citealt{Abazajian2009}) quasar catalog \citep{Shen2011}.
We chose \hbox{radio-quiet} quasars ($\rm R\_6cm\_2500\textup{\AA} <10$ in the \citealt{Shen2011} catalog) in the redshift range of $1.7 <z< 2.2$.
Radio-quiet quasars were chosen to avoid possible amplified (beamed) continua associated with relativistic jets, 
and the redshift range was chosen to have good coverage of the \iona{C}{iv} and \iona{Mg}{ii} line profiles in the SDSS spectra.
We excluded BAL quasars by requiring $\rm BAL\_FLAG = 0$ in the \citet{Shen2011} DR7 quasar catalog.
BAL quasars may have biased [\iona{C}{iv}] REW measurements due to the line absorption; they are also generally \xray\ absorbed \citep[e.g.,][]{Gallagher2002,Gallagher2006}.
For economical observations, we selected bright (\hbox{$\textit{i}$-band} magnitude $m_i \leq 18.2$) quasars.
We required the \iona{C}{iv} \hbox{emission-line} \hbox{signal-to-noise} ratio (SNR) $> 5$ ($\rm LINE\_MED\_SN\_CIV]>5$ in the \citealt{Shen2011} catalog).
Early studies employed the conservative REW $< 5$ -- $10~\textup{\AA}$ criterion to select \iona{C}{iv} WLQs \citep[e.g.,][]{Plotkin2010,Wu2012,Luo2015}.
However, \citet{Ni2018,Ni2022} constructed a representative sample of WLQs using a more inclusive criterion of REW $<15~\textup{\AA}$, which still show consistent multiwavelength properties.
Therefore, we adopted a threshold of \iona{C}{iv} REW $<15~\textup{\AA}$ for the initial selection.
We further supplemented our sample with the WLQs (29 unique WLQs after excluding objects in common) from \citet{Luo2015} that have \iona{C}{iv} measurements and \xray\ observations.
They have different redshift ($1.65<z<2.92$) and \hbox{$\textit{i}$-band} magnitude ($15.4\leq m_i \leq 18.5$) ranges from the SDSS DR7 selected targets.
Our observations were carried out with the Triple Spectrograph (TSpec) mounted on P200, which has spectral coverage of $\approx$0.95--2.46 $\rm \mu m$ and spectral resolutions of $\approx$2500--2700 \citep{Herter2008}.
We enforced an additional redshift filtering: $z<1.67$ and $2.08<z<2.65$, so that the broad $\rm H \beta$ emission line will not be significantly affected by telluric absorption.
We also excluded two objects with existing NIR spectra from \citet{Wu2011} or \citet{Plotkin2015}.
The parent sample constructed this way contains 28 WLQs.

We obtained 1.5 nights of good P200/TSpec observations in 2021 April and May via China's Telescope Access Program (TAP).\footnote{https://tap.china-vo.org/.}
We used a $1\arcsec \times 30\arcsec $ slit with standard ABBA nodding along the slit.
The exposure time for each target was either 40 or 60~minutes.
For flux calibration and telluric correction, we observed an A0V standard star before or after each observation.
The star was chosen so that the observation airmasses for the WLQ target and the star are comparable.
We obtained high SNR ($\gtrsim 10$) NIR spectra for 11 objects in the parent sample, which are all in the \citet{Shen2011} SDSS DR7 quasar catalog.
The basic information of these observations is presented in Table~\ref{tab:obslog}.
One target, SDSS~J$164302.03+441422.1$, turned out to have an incorrect redshift in the \citet{Shen2011} DR7 quasar catalog, and thus the WLQ identification was not reliable.
We removed this object from our sample; the spectrum of SDSS~J$164302.03+441422.1$ is presented in the Appendix.
The remaining ten objects constitute the final sample of this study.
Among these ten sources, nine have \iona{C}{iv} REW $<10~\textup{\AA}$ and four have \xray\ coverage from \cite{Wu2012}, \cite{Luo2015}, or \cite{Ni2022}. 
\begin{figure}
\centering \includegraphics[scale=0.35]{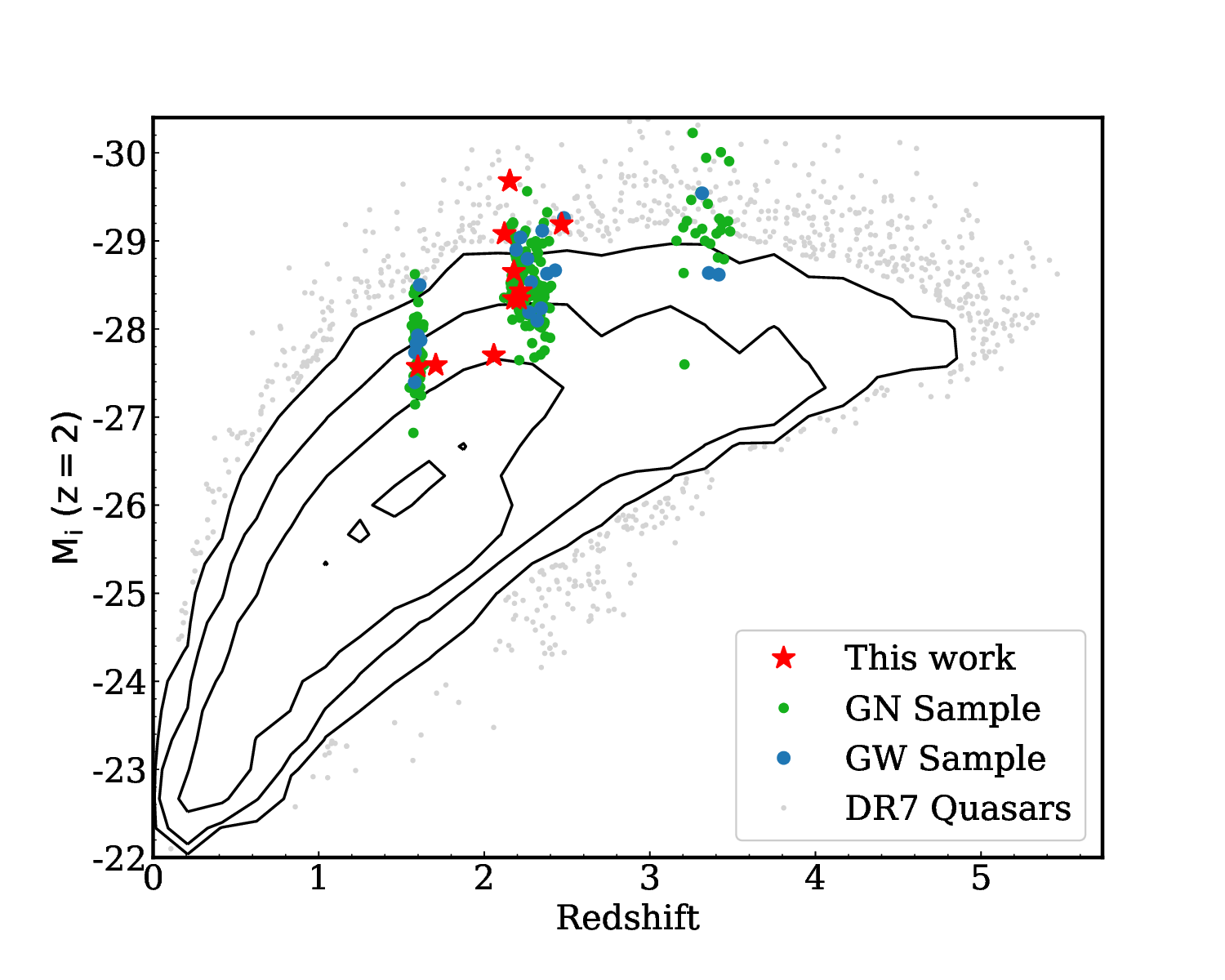}
\caption{Absolute \hbox{$\textit{i}$-band} magnitude vs.\ redshift for our ten objects (red stars). 
The underlying gray dots and black contours represent the 
distribution of objects in the SDSS DR7 quasar catalog.
The lowest contour level denotes $10^{1.6}$ quasars per bin (25 bins along each axis), 
and the increasing contours are logarithmically spaced by 0.6 dex; 
individual gray points are quasars falling below the lowest contour level.
Our sources are among the most luminous quasars.
The green and blue dots represent the GW and GN samples from the \citet{Matthews2023} \hbox{GNIRS-DQS} catalog that constitute our comparison samples.
}
\label{fig-mi_z}
\end{figure}
\subsection{Data Reduction}
\label{subsec:reduc}
\begin{deluxetable*}{lccccccc}
\tablecaption{P200/TSpec Observation Log}
\tablehead{
\colhead{Source Name}&
\colhead{$z_{\rm dr7}$$^{a}$}&
\colhead{$z_{\rm p200}$}&
\colhead{$i_{\rm psf}$$^{b}$}&
\colhead{$M_i$$^{c}$}&
\colhead{Obs. Date}&
\colhead{Exp. Time}&
\colhead{$\rm Ref_{X}$$^{d}$} \\
\colhead{(SDSS J)}&
\colhead{}&
\colhead{}&
\colhead{(mag)}&
\colhead{(mag)}&
\colhead{}&
\colhead{(min)}&
\colhead{}
}
\startdata
$073306.63+462517.5$ & 2.133 & 2.179 & 17.43 & $-28.65$ &  2021-04-02 & 40  &  --  \\
$085344.17+354104.5$ & 2.175 & 2.179 & 17.72 & $-28.34$ &  2021-04-02 & 40  &  --  \\
$111401.30+222211.4$ & 2.121 & 2.058 & 18.20 & $-27.70$ &  2021-04-02 & 40  &  --  \\
$122709.48+310749.3$ & 2.171 & 2.220 & 17.58 & $-28.43$ &  2021-04-02 & 40  &  --  \\
$134601.28+585820.2$ & 1.646 & 1.707 & 17.58 & $-27.59$ &  2021-05-21 & 60  &  (2), (3) \\
$153412.68+503405.3$ & 2.118 & 2.122 & 16.83 & $-29.08$ &  2021-04-02 & 40  &  (2)   \\
$153714.26+271611.6$ & 2.458 & 2.467 & 17.20 & $-29.19$ &  2021-04-02 & 40  &  (2)   \\
$154329.47+335908.7$ & 2.148 & 2.154 & 16.28 & $-29.68$ &  2021-05-21 & 60  &  --  \\
$154503.23+015614.7$ & 2.195 & 2.211 & 17.83 & $-28.34$ &  2021-05-21 & 60  &  --  \\
$161245.68+511816.9$ & 1.595 & 1.599 & 17.55 & $-27.57$ &  2021-05-21 & 60  &  (1)   \\
$164302.03+441422.1^{*}$ & 1.650 & 0.919 & 18.39 & $-26.81$ &  2021-05-21 & 60 & (2)
	\enddata
\tablenotetext{$\rm a$}{Redshift from \citet{Shen2011}.}
\tablenotetext{$\rm b$}{SDSS $i$-band point-spread function (PSF) magnitude from \citet{Schneider2010}.}
\tablenotetext{$\rm c$}{Absolute $i$-band magnitude ($K$-corrected to $z=2$) from \citet{Shen2011}. }
\tablenotetext{$\rm d$}{Reference papers for the objects with published \xray\ data. (1): \citet{Wu2012}; (2): \citet{Luo2015}; (3): \citet{Ni2022}.}
\tablenotetext{*}{This object has an incorrect redshift in the \citet{Shen2011} catalog.
It is a WLQ candidate and it is not included in our final sample. }
\label{tab:obslog}	
\end{deluxetable*}

We reduced raw P200/TSpec observational data following the standard procedure, described briefly below.
The modified IDL-based Spextool~4 package \citep{Cushing2014} was used for data reduction.
We first constructed dark and flat-field files.
For each source, wavelength calibration was then performed based on automatic identification of telluric airglow lines in the spectra of the corresponding standard star.
We extracted one-dimensional spectra with the ``A$-$B'' method to subtract background.
Each source thus has 8 (for 40-minute exposures) or 12 (for 60-minute exposures) spectra and each spectrum consists of four orders that are to be merged in the end.
We combined the spectra for each object 
by scaling them to the spectrum that has the highest flux level in the $\rm H\beta$ region (\hbox{rest-frame} 4360--5360 $\textup{\AA}$) and then stacking all the spectra.
The brightest spectrum is likely the least affected by guiding errors and variable sky transmission, and thus it is considered the most accurate spectrum for flux normalization.

We then corrected telluric absorption following the procedure described in \citet{Vacca2003}.
We first extracted spectra of the corresponding standard star with the same approach above.
Given the $B$- and $V$-band magnitudes of the star, we created a model spectrum of Vega,
convolved to the spectral resolution of the combined star spectra.
The model spectrum was then divided by the star spectrum to yield a telluric correction spectrum.
We visually inspected each telluric correction spectrum and verified that there are no significant residual hydrogen lines.
We multiplied the object spectrum by the telluric correction spectrum to produce a flux-calibrated and telluric-corrected spectrum.
Finally, we merged the four orders in each spectrum by normalizing them using the overlapping regions;
typically the scaling factors are close to unity ($\approx$ 0.97--1.05) between the orders.

We examined the quality of the final spectra.
The average SNRs per pixel in the H$\beta$ region for our ten sources range from $\approx 10$ to $\approx 50$ (with a median value of $\approx 17$), sufficient for basic spectral analysis.
We also compared our spectra to the available SDSS spectra by dividing them in the overlapping regions.
The average ratios range from 0.75 to 1.42 with a median value of 0.95.
These flux uncertainties are likely dominated by systematic uncertainties in the above flux calibration procedure, with additional contributions from the SDSS flux calibration uncertainties and quasar flux variability over timescales of years  \citep[e.g.,][]{Shen2021,Suberlak2021}.

\subsection{Spectral Analysis}
\label{subsec:spec}
\begin{figure*}[h!]
\centering \includegraphics[scale=0.30]{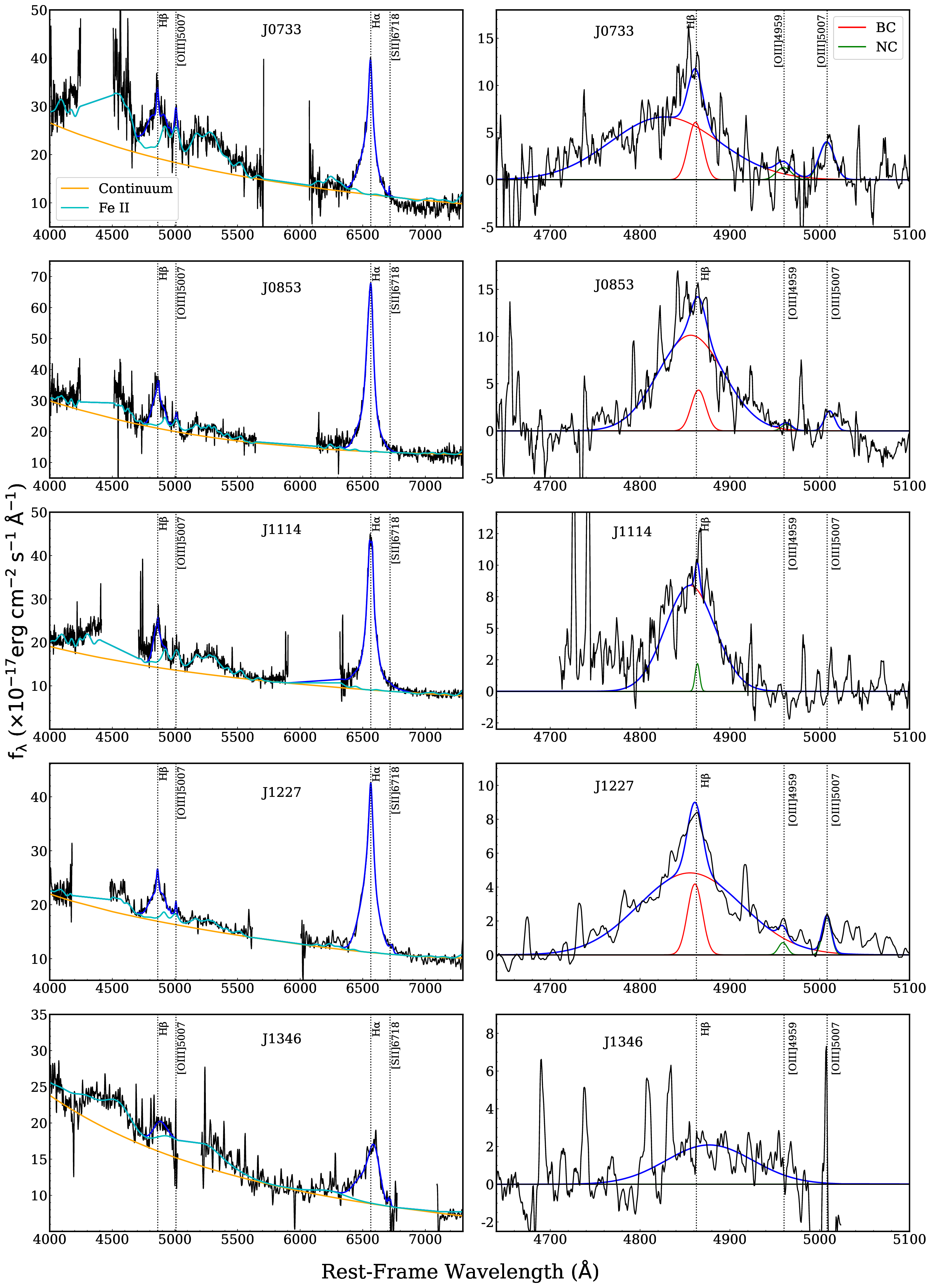}
\caption{
\hbox{Rest-Frame} optical spectra for our ten sources and the \hbox{best-fit} models.
Positions of major emission lines are marked.
The left panel shows the broad-band spectra (black curves)
smoothed with box convolutions (ranging from 5 to 10 pixels).
The blue curve represents the \hbox{best-fit} continuum $+$ \hbox{emission-line} model,
The orange curve shows the \hbox{power-law} $+$ polynomial (if present) continuum,
and the cyan curve shows the \iona{Fe}{ii} pseudo continuum.
The right panel
shows zoomed-in views of the spectra in the H$\rm \beta$ region, 
fitted with broad (red curves) 
and/or narrow (green curves) Gaussian profiles 
for the H$\rm \beta$ and [\iona{O}{iii}] emission lines.
}
\label{fig-spec1}
\end{figure*}
\renewcommand{\thefigure}{\arabic{figure} (Cont.)}
\addtocounter{figure}{-1}
\begin{figure*}
\centering \includegraphics[scale=0.30]{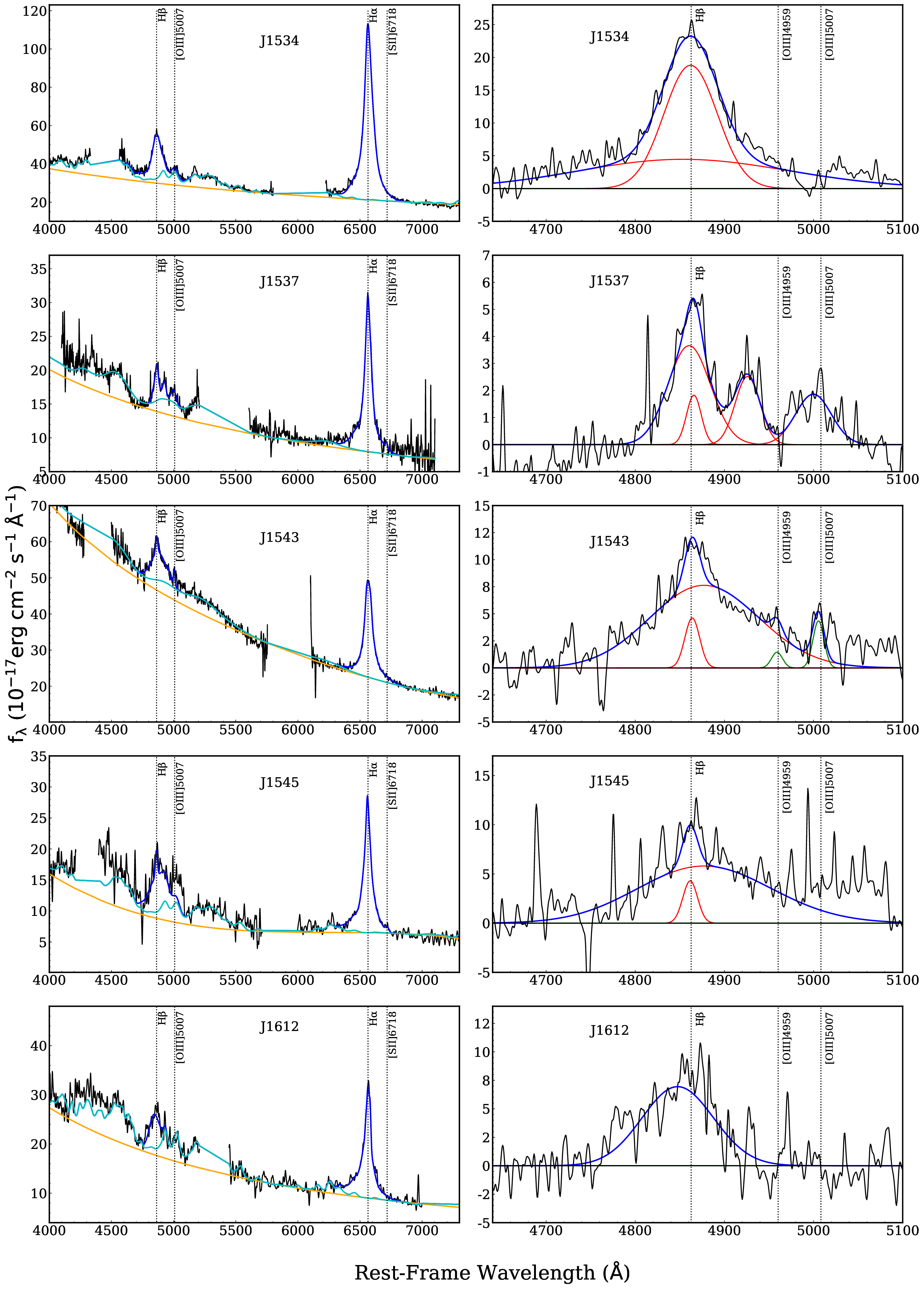}
\caption{}
\end{figure*}
\renewcommand{\thefigure}{\arabic{figure}}
Spectral analyses of the P200/TSpec spectra were performed with the PyQSOFit package (v1.1),\footnote{https://github.com/legolason/PyQSOFit.} which uses a 
\hbox{$\chi^{2}$-based} method to fit the continuum and emission lines \citep[e.g.,][]{Shen2011,pyqsofit,Guo2019}.
PyQSOFit first applied Galactic extinction corrections to the spectra using the $E(B-V)$ values from \citet{Schlegel1998} and the Galactic extinction curve from \citet{Fitzpatrick1999} with $R_V = 3.1$.
Our continuum model
consists of a \hbox{\hbox{power-law}} component, a polynomial component, the \citet{Boroson1992} optical \iona{Fe}{ii} emission
template, and a \hbox{Balmer-continuum} template. The Balmer-continuum component is not needed for six of the ten sources.
The polynomial component is used to account for spectral complexity such as reddening, and it is not needed for 
four sources. The fitting was performed using line-free windows around major broad emission lines (e.g., 6000--6250~$\textup{\AA}$ and 6800--7000~$\textup{\AA}$
around H$\alpha$, 4435--4630~$\textup{\AA}$ and 5100--5535~$\textup{\AA}$ around H$\beta$).
Given the \hbox{best-fit} continuum, we derived the monochromatic luminosity at 5100~$\textup{\AA}$ ($L_{\rm 5100}$) from the \hbox{power-law} component
$+$ the polynomial component (if present), and we measured the REW of the \iona{Fe}{ii}
pseudo continuum between 4434 and 4684~$\textup{\AA}$ against the underlying continuum.

After subtracting the \hbox{best-fit} continuum, we fitted the emission lines with multiple Gaussian profiles. 
All line fits were performed in the rest frame. 
For AGN broad emission-lines, a single Gaussian function usually cannot explain the line profile, and multiple Gaussian profiles (sometimes with additional Lorentzian profiles) are often employed to fit the lines \citep[e.g.,][]{Shen2011,Perna2015,Plotkin2015,DU2019,Guo2019}.
The reason behind this is complex.
It might be due to a combination of line asymmetries from absorption or outflows, broad-line stratification, and inaccurate subtraction of the continuum (including the \iona{Fe}{ii} pseudo continuum).
Because of these effects, the multiple Gaussian profiles used to fit the lines do not carry individual physical significance.
For example, the velocity offsets of individual components usually do not represent outflow velocities.
The line parameters (e.g., REW, FWHM, flux) are thus typically measured using the combined line profile.

For the H$\beta$ line, we used up to three Gaussians, including two broad components with full widths at half maximum (FWHMs)~$\rm \geq 1200~km~s^{-1}$
and one narrow component with FWHM~$\rm < 1200~km~s^{-1}$; 
this $\rm 1200~km~s^{-1}$ \hbox{line-width} criterion also applies to other lines in this study.
Among the ten sources, seven require two broad components and the other three have only one broad component. 
Only one source shows a clear narrow H$\beta$ component.
Following \citet{Shen2011}, we adopted a \hbox{Monte-Carlo} approach to obtain uncertainties of the \hbox{best-fit} parameters.
For each source, 1000 mock spectra were created by randomizing the spectrum with the Gaussian noise at individual pixels.
We then fitted the mock spectra following the same procedure above, and the 
1$\sigma$ dispersion of the 1000 measurements for each parameter was adopted as its uncertainty.

We used two narrow Gaussians and two broad Gaussians to fit the [\iona{O}{iii}]~$\lambda 4959, 5007$ doublet; 
we fixed the flux ratio of the narrow components to the theoretical value of 3.0 (i.e., $f_{5007}/f_{4959} = 3$).
We tied the velocity offsets of the narrow H$\beta$ line and the [\iona{O}{iii}]~$\lambda 4959, 5007$ doublet if they are available.
Some sources do not show clear [\iona{O}{iii}] lines.
We consider the [\iona{O}{iii}]~$\lambda 5007$ line to be detected if the \hbox{best-fit} REW exceeds three times the 1$\sigma$ dispersion determined from the Monte Carlo approach above.
Four sources have narrow [\iona{O}{iii}]~$\lambda 5007$ REW measurements and another object, SDSS J153714.26 $+$ 271611.6, appears to have broad [\iona{O}{iii}] components.
The broad components have significant blueshifts and might indicate large-scale ionized outflows in this quasar (see Section~\ref{subsec:outflow} below), and
thus we do not consider the broad components part of typical AGN [\iona{O}{iii}] emission.
For the six objects without narrow [\iona{O}{iii}] lines, we present their 3$\sigma$ upper limits on the [\iona{O}{iii}]~$\lambda 5007$ REWs;
such stringent upper limits were adopted mainly for the survival analysis in Section~\ref{subsec:elp} below.

For the H$\alpha$ complex,
we adopted up to three broad Gaussians and one narrow Gaussian for the H$\alpha$ line,
and we used up to four narrow Gaussians to
model the [\iona{N}{ii}] and [\iona{S}{ii}] lines.
We enforced a fixed flux ratio for the narrow [\iona{N}{ii}] lines by setting $f_{6584}/ f_{6548} = 3$ \citep[e.g.,][]{Shen2011}, as the lines are typically located on top of the broad H$\alpha$ profile core.
We tied the velocity offsets of the narrow H$\alpha$ line and the [\iona{N}{ii}] and [\iona{S}{ii}] line sets.
However, we did not tie these velocity offsets to those of the narrow H$\beta$ and [\iona{O}{iii}] lines as the latter lines are often weak or absent.
Basic \hbox{emission-line} parameters, including REW, FWHM, and line flux,
were derived from the combined broad or narrow components for each line.

We improved the redshifts of our sources using the P200/TSpec spectra. 
Their redshifts from the SDSS DR7 quasar catalog \citep{Shen2011}
are listed in Table~\ref{tab:obslog}, which are consistent with those in \citet{Hewett2010}.
These redshifts were derived based on mainly the weak and sometimes blueshifted 
UV emission lines (e.g., \iona{C}{iv} and \iona{Mg}{ii}).
Since the majority of our sources do not show clear [\iona{O}{iii}] line emission in the P200/TSpec spectra, 
we obtained their redshifts from the peak positions of the broad H$\alpha$ and H$\beta$ \hbox{emission-line} profiles adopting an iterative procedure. 
The \citet{Shen2011} redshift was used as the initial input in PyQSOFit. 
We then measured a new redshift from the average velocity offset of the \hbox{best-fit} H$\alpha$ and H$\beta$ broad components. 
The new redshift was used in the next iteration. 
If the redshift difference is smaller than 0.001, we considered the value converged 
and stopped the iteration. 
The resulting redshifts are listed in Table~\ref{tab:obslog}, and they differ slightly 
(in the range of $-0.063\textrm{--}0.061$ with an average value of $-0.014$)
from the \citet{Shen2011} redshifts for our ten sources.
We adopt the P200/TSpec redshifts in the following analyses.

We show the absolute $i$-band magnitude versus redshift distribution for our sources in Figure~\ref{fig-mi_z}; these are luminous quasars compared to
typical SDSS DR7 quasars due to our selection of bright targets.
The spectra and the 
\hbox{best-fit} models are displayed in Figure~\ref{fig-spec1}, and the \hbox{emission-line} measurements are summarized in Table~\ref{tab:elp}.

\begin{deluxetable*}{lcccccccc}
\tablecaption{Emission-Line Properties}
\tablehead{
\colhead{Source Name}&
\colhead{$\rm H \beta$~REW}&
\colhead{[\iona{O}{iii}]~$\lambda 5007$~REW$^{a}$}&
\colhead{\iona{Fe}{ii}~REW}&
\colhead{$\rm H \beta$~FWHM}&
\colhead{\iona{Fe}{ii}~FWHM}&
\colhead{\iona{C}{iv}~REW$^{c}$}& \\
\colhead{(SDSS J)}&
\colhead{($\rm \textup{\AA}$)}&
\colhead{($\rm \textup{\AA}$)}&
\colhead{($\rm \textup{\AA}$)}&
\colhead{($\rm km~s^{-1}$)}&
\colhead{($\rm km~s^{-1}$)}&
\colhead{($\rm \textup{\AA}$)}& \\
}
\startdata
$073306.63+462517.5$ &  $58.8\pm1.4$&   $4.5\pm0.6$   & $101\pm36$  & $4900\pm280$&   $3290\pm2100$& $6.4\pm1.2$   \\
$085344.17+354104.5$ &  $50.7\pm0.7$&   $1.7\pm0.4$   & $36\pm2$   &  $3780\pm210$&   $1740\pm80$&   $3.5\pm0.8$   \\
$111401.30+222211.4$ &  $55.4\pm1.4$&   $<0.98$      & $79\pm9$   &  $4450\pm310$&   $2410\pm140$&   $3.9\pm0.9$   \\
$122709.48+310749.3$ &  $45.1\pm1.4$&   $1.7\pm0.2$ & $30\pm4$   &  $2700\pm270$&   $2420\pm170$&    $13.5\pm1.1$  \\
$134601.28+585820.2$ &  $17.3\pm1.3$&   $<0.91$      & $56\pm5$   &  $6990\pm1100$&  $10000^{b}$&    $2.4\pm0.9$   \\
$153412.68+503405.3$ &  $91.1\pm1.5$&   $<3.3$      & $59\pm3$   &  $5040\pm65$&    $2390\pm600$&    $9.2\pm0.9$   \\
$153714.26+271611.6$ &  $24.9\pm0.4$&   $<3.1$      & $50\pm2$   &  $2500\pm1200$&  $4870\pm250$&    $7.9\pm0.7$   \\
$154329.47+335908.7$ &  $25.7\pm0.4$&   $1.5\pm0.3$ & $25\pm1$   &  $5090\pm190$&   $10000^{b}$&      $8.1\pm0.5$   \\
$154503.23+015614.7$ &  $134.6\pm4.8$&  $<13.1$      & $88\pm11$  &  $5130\pm1400$&  $3150\pm260$&    $9.6\pm0.9$   \\
$161245.68+511816.9$ &  $25.4\pm1.4$&   $<2.0$      & $70\pm4$   &  $4870\pm360$&   $2370\pm360$&    $1.6\pm1.9$ 
\enddata
\tablenotetext{$\rm a$}{Measurement or upper limit for the narrow [\iona{O}{iii}]~$\lambda 5007$ REW. SDSS J153714.26 $+$ 271611.6 appears to have broad [\iona{O}{iii}] components (see Section~\ref{subsec:outflow} below).}
\tablenotetext{$\rm b$}{This FWHM value pegged at $\rm 10000~km~s^{-1}$, the
upper bound allowed by our PyQSOFit fitting.}
\tablenotetext{$\rm c$}{The \iona{C}{iv} REW is adopted from \citet{Shen2011}.}
\label{tab:elp}
\end{deluxetable*}

\subsection{\texorpdfstring{High-$z$}{} Comparison Samples}
\label{subsec:comp}

To identify distinct features in the NIR spectra of the WLQs in our sample, we selected comparison samples that consist of high-redshift quasars with NIR spectra.
These were drawn from the 260 quasars in the Gemini Near Infrared Spectrograph-Distant Quasar Survey (\hbox{GNIRS-DQS}; \citealt{Matthews2021,Matthews2023}),
which is a homogeneous flux-limited NIR spectroscopic survey of high-redshift quasars.
We first chose the 199 \hbox{GNIRS-DQS} quasars that are also in the \citet{Shen2011} SDSS DR 7 quasar catalog in order to have good \iona{C}{iv} REW measurements ($\rm LINE\_MED\_SN\_CIV]>5$) from that catalog, consistent with our target selection.
We then excluded significantly \hbox{radio-loud} ($R>100$)\footnote{The \hbox{radio-loudness} parameter ($R$) is defined as $R=f_{5~{\rm GHz}}/f_{\rm 4400~{\textup{\AA}}}$ \citep{Kellermann1989}, where $f_{5~{\rm GHz}}$ and $f_{\rm 4400~{\textup{\AA}}}$ are the flux densities at 5~GHz and $\rm 4400~{\textup{\AA}}$, respectively.}
quasars from this sample
as the optical continua of these quasars might have potential contamination from \hbox{non-thermal} jet radiation. 
To obtain the \hbox{radio-loudness} parameters, we matched the \hbox{GNIRS-DQS} objects to the catalogs of the 
Faint Images of the Radio Sky at \hbox{Twenty-cm} (FIRST; \citealt{Becker1995}) survey and the NRAO Very Large Array (VLA) Sky Survey (NVSS; \citealt{Condon1998}) with matching radii of $5\arcsec$ and $30\arcsec$, respectively.
A radio \hbox{power-law} spectral slope of $\alpha_{r} = -0.8$ was adopted to convert 1.4~GHz flux densities to 5~GHz flux densities. 
Nine \hbox{GNIRS-DQS} quasars, all detected in FIRST, have $R>100$
and they were excluded from the comparison
samples, leaving 190 \hbox{GNIRS-DQS} quasars.

There are also WLQs in the \hbox{GNIRS-DQS} sample.
Adopting the same
\iona{C}{iv} REW $< 15$~$\textup{\AA}$ criterion using the SDSS DR7 measurements, 
we identified 22
WLQs (12 having REW $< 10$~$\textup{\AA}$), two of which are in common with our objects (SDSS~J$085344.17+354104.5$ and
SDSS~J$122709.48+310749.3$).
After excluding these WLQs, we refer to the remaining 168 ($190-22$) quasars 
as the Gemini-normal (GN) sample which is our comparison sample of 
typical quasars.
For the 20 \hbox{GNIRS-DQS} WLQs (excluding the two objects in common),
the measurements were obtained from a different instrument with a different spectral analysis approach.
We thus do not consider it appropriate to combine them with our objects to construct a larger sample.
Instead, we consider these 20 WLQs the Gemini-weak (GW) sample,
and also compare our sample properties to theirs.
The absolute $i$-band magnitude versus redshift distributions for the two comparison samples are also shown in Figure~\ref{fig-mi_z}.
Except for the z > 3 objects, they occupy similar regions as the WLQs in our sample.
We have verified that excluding the 24 (3) z > 3 quasars from the GN (GW) sample does not change our comparison results in Section~\ref{subsec:elp} below.

Measurements of the H$\beta$, [\iona{O}{iii}], and \iona{Fe}{ii} emission lines for the two comparison samples are adopted from \citet{Matthews2023}.
As a consistency check, we randomly selected ten quasars in the comparison samples and measured their \hbox{emission-line} properties using the \hbox{GNIRS-DQS} spectra, 
and the results are overall consistent with those in \citet{Matthews2023}. 
For the two WLQs in common, SDSS~J$085344.17+354104.5$ and SDSS~J$122709.48+310749.3$, 
we also compared their P200 and \hbox{GNIRS-DQS} measurements. 
The \hbox{emission-line} properties for SDSS~J$085344.17+354104.5$ are consistent within the errors, 
but they differ by up to $\approx50\%$ for SDSS~J$122709.48+310749.3$,
which are likely attributed to systematic uncertainties from the flux calibration and spectral analysis approaches.
We do not observe any systematic bias between our data analysis procedure and that in \citet{Matthews2023}, and thus statistical comparisons of the sample properties should guard against such uncertainties.
\citet{Matthews2023} have a different [\iona{O}{iii}] detection threshold for the GNIRS-DQS quasars and objects with [\iona{O}{iii}] REWs below $1~\textup{\AA}$ were considered non-detections. 
We set their REW upper limits to $5~\textup{\AA}$ following \citet{Matthews2021}.
As objects with [\iona{O}{iii}] non-detections constitute a small fraction (10\%) in the GN sample and a large fraction (45\%) in the GW sample, the exact upper-limit value for the non-detections does not change qualitatively our statistical comparison results in Section~\ref{subsec:elp} below.

\section{Results}
\label{sec:result}

\subsection{Emission-Line Properties and Comparisons}
\label{subsec:elp}
\begin{figure*}
\centering \includegraphics[scale=0.35]{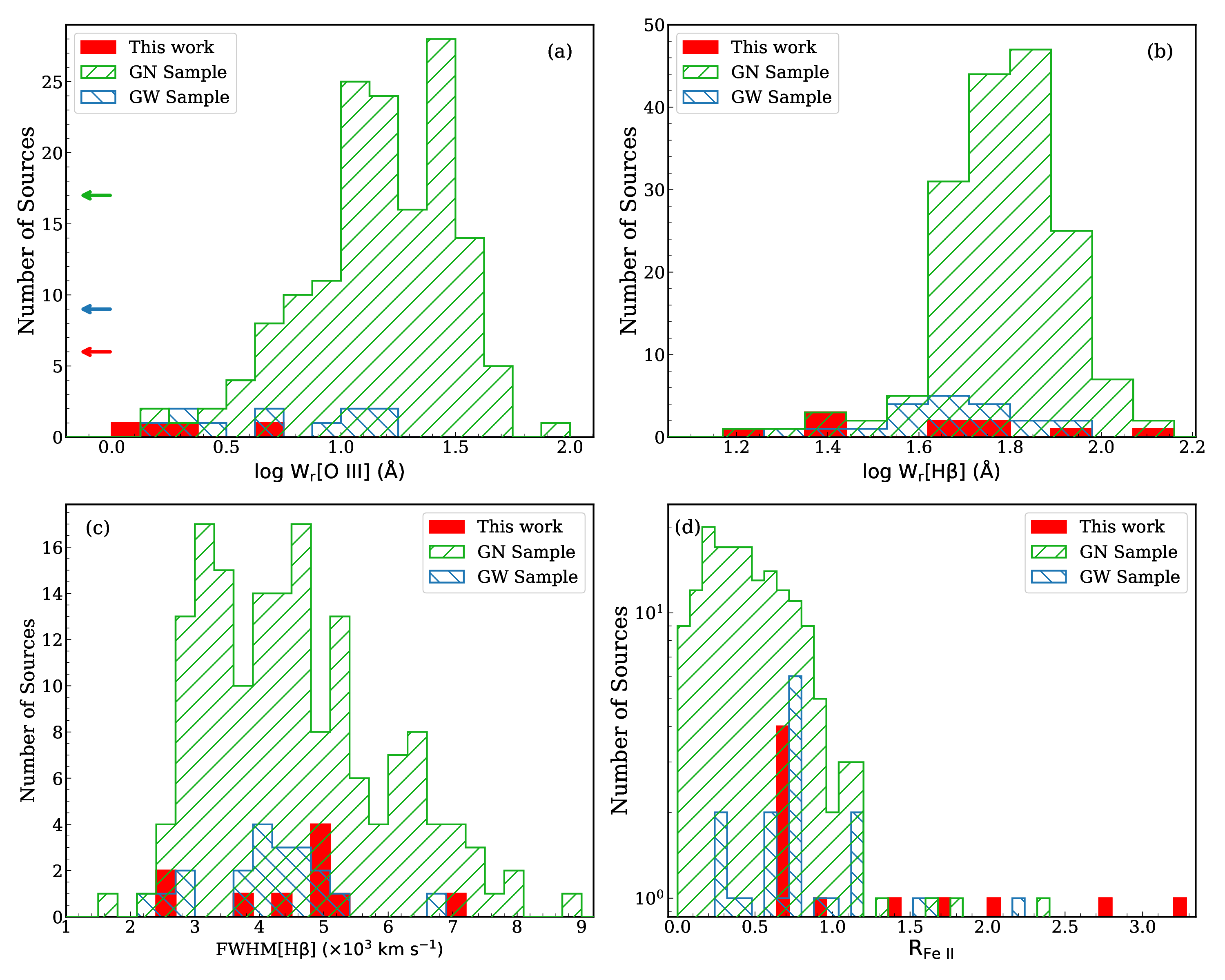}

\caption{Distributions of (a) log $W_{r}$[\iona{O}{iii}], (b) log $W_{r}$[H$\rm \beta$], (c) H$\rm \beta$ FWHMs, and (d) $R_{\rm Fe\ II}$ for our WLQ sample. 
For comparison, the distributions for the GW and GN samples are also shown in each panel. 
For the $W_{r}$[\iona{O}{iii}] distributions, upper limits are indicated by the leftward arrows and the y-axis values of the arrows represent the numbers of sources with upper limits in the corresponding samples.
Our WLQs show weak $W_{r}$[\iona{O}{iii}] and stronger \iona{Fe}{ii} emission compared to typical quasars. 
Their $\rm H \beta$ emission is not particularly weak (K-S test $P_{\rm null}=0.06$).}
\label{fig-elp}
\end{figure*}

The main aim of this study is to examine the \hbox{rest-frame} optical spectroscopic properties of WLQs.
The \hbox{best-fit} continuum shapes of our objects, determined from the \hbox{power-law} component $+$ the polynomial component (if present; Section~\ref{subsec:spec}), appear to be similar to those of typical SDSS quasars \citep[e.g.,][]{Vanden2001}.
Nominal optical continuum shapes were also reported for the WLQ sample in \citet{Plotkin2015}.
These are consistent with the typical quasar IR--UV continuum SEDs observed for WLQs \citep[e.g.,][]{Lane2011,Luo2015,Ni2018}.
Therefore, we focus below on the investigation of \hbox{emission-line} properties.

\begin{deluxetable}{lccc}
\tablecaption{Statistical Test Results for the Emission-Line Properties of Our Sample and the Comparison Samples}
\tablehead{
\colhead{}&
\colhead{}&
\multicolumn{2}{c}{Peto-Peto Generalized Wilcoxon} \\
\colhead{Parameter}&
\colhead{Test samples}&
\multicolumn{2}{c}{or Kolmogorov-Smirnov Test$^{a}$} \\
\cline{3-4} \\
\colhead{} &
\colhead{} & 
\colhead{Statistic} &  
\colhead{$P_{\rm null}$}   
}
\startdata
$W_{r}$[\iona{O}{iii}]      &P200 vs. GN   & 4.70 & $3\times 10^{-6}$  \\
							&P200 vs. GW   & 2.69 & 0.007 \\
							&&&\\
$W_{r}$[H$\beta$]           &P200 vs. GN   & 0.41 & 0.06  \\
							&P200 vs. GW   & 0.30 & 0.57   \\
							&&&\\ 
FWHM[H$\beta$]              &P200 vs. GN   & 0.24 & 0.58 \\
							&P200 vs. GW   & 0.40 & 0.22 \\
							&&&\\
$W_{r}$[\iona{Fe}{ii}]      &P200 vs. GN   & 0.59 & 0.001 \\
							&P200 vs. GW   & 0.50 & 0.06  \\
							&&&\\                           
$R_{\rm Fe\ II}$            &P200 vs. GN   & 0.71 & $3\times10^{-5}$ \\
							&P200 vs. GW   & 0.35 & 0.37
	\enddata
\label{tab:Samplestats}
\tablenotetext{$\rm a$}{The Peto-Peto Generalized Wilcoxon test was run for the $W_{r}$[\iona{O}{iii}] comparison, and the Kolmogorov-Smirnov tests were run for the other comparisons.}
\end{deluxetable}	

The \hbox{emission-line} features we examined include the [\iona{O}{iii}]~$\lambda 5007$ REW ($W_{r}$[\iona{O}{iii}]), H$\beta$ REW ($W_{r}$[H$\beta$]), H$\beta$ FWHM (FWHM[H$\beta$]), optical \iona{Fe}{ii} REW ($W_{r}$[\iona{Fe}{ii}]), and ${R}_{\rm Fe\ II}$ ($W_{r}$[\iona{Fe}{ii}]/$W_{r}$[H$\beta$]).
For statistical comparisons to the comparison samples, we ran the Peto-Peto Generalized Wilcoxon (P-P) test in the Astronomy Survival Analysis package (ASURV; e.g., \citealt{Feigelson1985,Lavalley1992}) 
for the $W_{r}$[\iona{O}{iii}] measurements which contain censored data, and we ran the Kolmogorov-Smirnov (K-S) test for the other quantities.
The test results (test statistics and null-hypothesis probabilities) are listed in Table~\ref{tab:Samplestats}.
The distributions of $W_{r}$[\iona{O}{iii}], $W_{r}$[H$\beta$], FWHM[H$\beta$] and ${R}_{\rm Fe\ II}$ for our sample and the comparison samples are displayed in Figure~\ref{fig-elp}.

With only four objects having measurable [\iona{O}{iii}] emission lines (Table~\ref{tab:elp}), the WLQs in our sample show noticeably weak [\iona{O}{iii}] emission compared to the GN sample ($P_{\rm null}=3\times 10^{-6}$). 
There appears to be a small difference between the $W_{r}$[\iona{O}{iii}] distributions of our sample and the GW sample ($P_{\rm null}=0.007$).
We ran a ASURV P-P test on the GW versus GN sample, and they show significantly different $W_{r}$[\iona{O}{iii}] distributions ($P_{\rm null}=2\times10^{-6}$).
Thus, the GW quasars still show weaker [\iona{O}{iii}] emission than typical quasars.
As one of the Eigenvector~1 parameters, $W_{r}$[\iona{O}{iii}] has been suggested to be negatively correlated with the accretion rate \citep[e.g.,][]{Boroson1992,Sulentic2000,Shen2014}. 
Thus, it is not surprising to observe WLQs with  weak [\iona{O}{iii}] emission under the scenario that these are \hbox{super-Eddington} accreting with soft ionizing continua from TDO obscuration. 
In the literature, both of the two WLQs in \citet{Shemmer2010} show weak [\iona{O}{iii}] emission. 
Only two objects in the WLQ sample of \cite{Plotkin2015} have good [\iona{O}{iii}] coverage, and they both show weak [\iona{O}{iii}] emission.
In \citet{Ha2023}, the two new WLQs with NIR spectroscopy (see their Appendix) do not show any detectable [\iona{O}{iii}] emission lines.
PHL~1811, a well-studied luminous WLQ at $z = 0.192$, lacks [\iona{O}{iii}] emission in its optical spectrum \citep[e.g.,][]{Leighly2007}. 
J1521$+$5202, the only PHL~1811 analog in \citet{Wu2011} with an NIR spectrum, does not show appearent [\iona{O}{iii}] emission either.
A recent James Webb Space Telescope (JWST) Near-InfraRed Camera (NIRCam) spectrum of the $z= 6.327$ WLQ, J0100$+$2802 \citep{Wuxb2015}, shows no narrow [\iona{O}{iii}] emission either \citep{Eilers2022}.

Weak [\iona{O}{iii}] emission has also been observed in other high-$z$ luminous quasar samples. 
For example, $\approx 70\%$ of the WISE/SDSS selected hyper-luminous (WISSH) quasars show weak or absent [\iona{O}{iii}] emission; several of these also show weak \xray\ emission and a few have \iona{C}{iv} REW $< 15~\textup{\AA}$ and could be considered WLQs \citep[e.g.,][]{Vietri2018,Zappacosta2020}. 

Our WLQ sample does not show significantly weaker H$\beta$ emission (average $W_{r}$[H$\beta$] $\rm \approx 52.9\pm2.0~\textup{\AA}$) compared to the GN sample (average $W_{r}$[H$\beta$] $\rm \approx 64.4\pm8.1~\textup{\AA}$). 
The K-S test also suggests that the two H$\beta$ REW distributions do not differ much ($P_{\rm null}=0.06$). 
The H$\beta$ REW distributions for our sample and the GW sample are similar ($P_{\rm null}=0.57$). 
\citet{Plotkin2015} found that the WLQ H$\beta$ lines are toward the weaker end of the distribution for typical quasars, but they are not as exceptionally weak as for the \hbox{high-ionization} \iona{C}{iv} lines. 
For the H$\beta$ FWHMs, there is no significant difference between our sample and the comparison samples; 
the average H$\beta$ FWHMs for our WLQ sample is $\rm 4540 \pm 710~km~s^{-1}$. 
It has been suggested that the FWHMs of the H$\beta$ lines have an orientation dependence \citep[e.g.,][]{Wills1986,Runnoe2013,Shen2014}, and larger FWHMs generally correspond to larger inclination angles. 
WLQs probably do not have unusual inclination angles compared to typical quasars at similar redshifts, although it may be interesting to compare the H$\beta$ FWHMs between \xray\ weak and \xray\ normal WLQs (see discussion in Section~\ref{subsec:scenario} below).

\begin{figure}[h!]
\centering \includegraphics[scale=0.35]{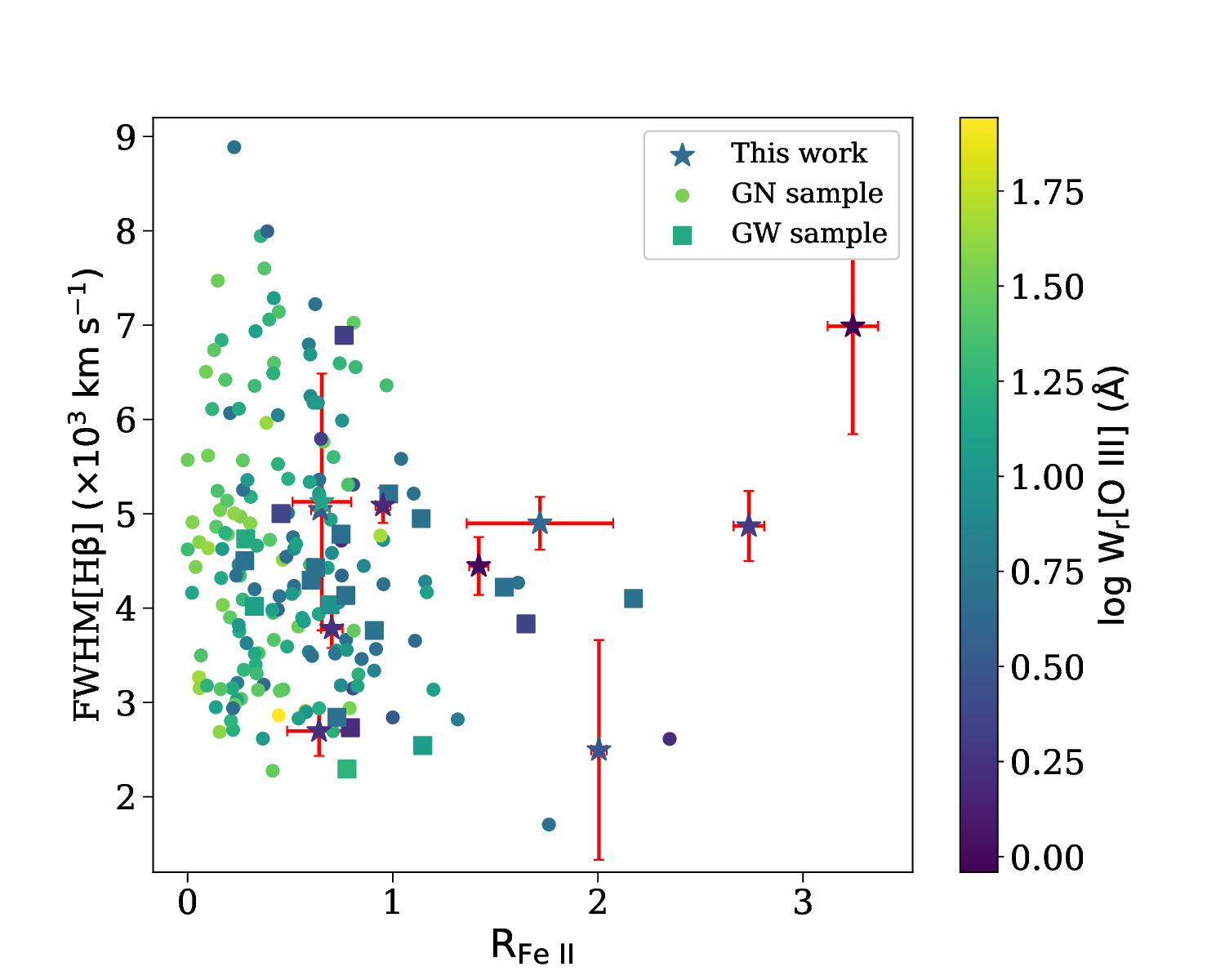}

\caption{FWHM[H$\beta$] vs. $R_{\rm Fe\ II}$ for our sample (stars), the GN sample (dots), and the GW sample (squares).
We color-code the points by $W_{r}$[\iona{O}{iii}] to illustrate the distribution of these quasars in the Eigenvector~1 plane.
The error bars for the comparison samples are not shown for display purposes.
Compared to typical quasars, WLQs show stronger optical \iona{Fe}{ii} emission and weaker [\iona{O}{iii}] emission in general.}
\label{fig-fwhmrfe}
\end{figure}

Stronger optical \iona{Fe}{ii} emission appears to be another distinct feature of WLQs. 
The probability that the $W_{r}$[\iona{Fe}{ii}] distributions of our WLQ sample and the GN sample are drawn from the same parent population is only $P_{\rm null}=0.001$.
The difference is enhanced ($P_{\rm null}=3\times 10^{-5}$) if ${R}_{\rm Fe\ II}$ is considered due to the slightly weaker H$\beta$ emission for WLQs.
The $W_{r}$[\iona{Fe}{ii}] and ${R}_{\rm Fe\ II}$ distributions for our sample and the GW sample are similar ($P_{\rm null}=0.06$ and $P_{\rm null}=0.37$, respectively).
Similar findings have been reported for the WLQs in \citet{Shemmer2010} and \citet{Plotkin2015}. 
Being another important Eigenvector~1 parameter, ${R}_{\rm Fe\ II}$ is negatively correlated with $W_{r}$[\iona{O}{iii}], and large ${R}_{\rm Fe\ II}$ values are likely also driven by high accretion rates \citep[e.g.,][]{Boroson1992,Sulentic2000,Shen2014}, consistent with the \hbox{super-Eddington} scenario for WLQs. 
Following \citet{Shen2014}, we plot in Figure~\ref{fig-fwhmrfe} the FWHM[H$\beta$] versus ${R}_{\rm Fe\ II}$ distributions with color-coded $W_{r}$[\iona{O}{iii}] for our sample and the comparison samples. In general, the WLQs show weaker [\iona{O}{iii}] emission, enhanced \iona{Fe}{ii} emission, and typical H$\beta$ FWHM values.
A few objects (e.g., SDSS~J$134601.28+585820.2$, the rightmost data point) appear to be outliers in the FWHM[H$\beta$] versus ${R}_{\rm Fe\ II}$ correlation for typical quasars \citep[e.g.,][]{Boroson1992,Shen2014}, but these could be explained with object-to-object intrinsic scatter of the correlation or systematic uncertainties from the spectral analysis.

\subsection{SMBH Mass and Eddington-Ratio Estimates}
\label{subsec:bh}
\begin{figure}
\centering \includegraphics[scale=0.35]{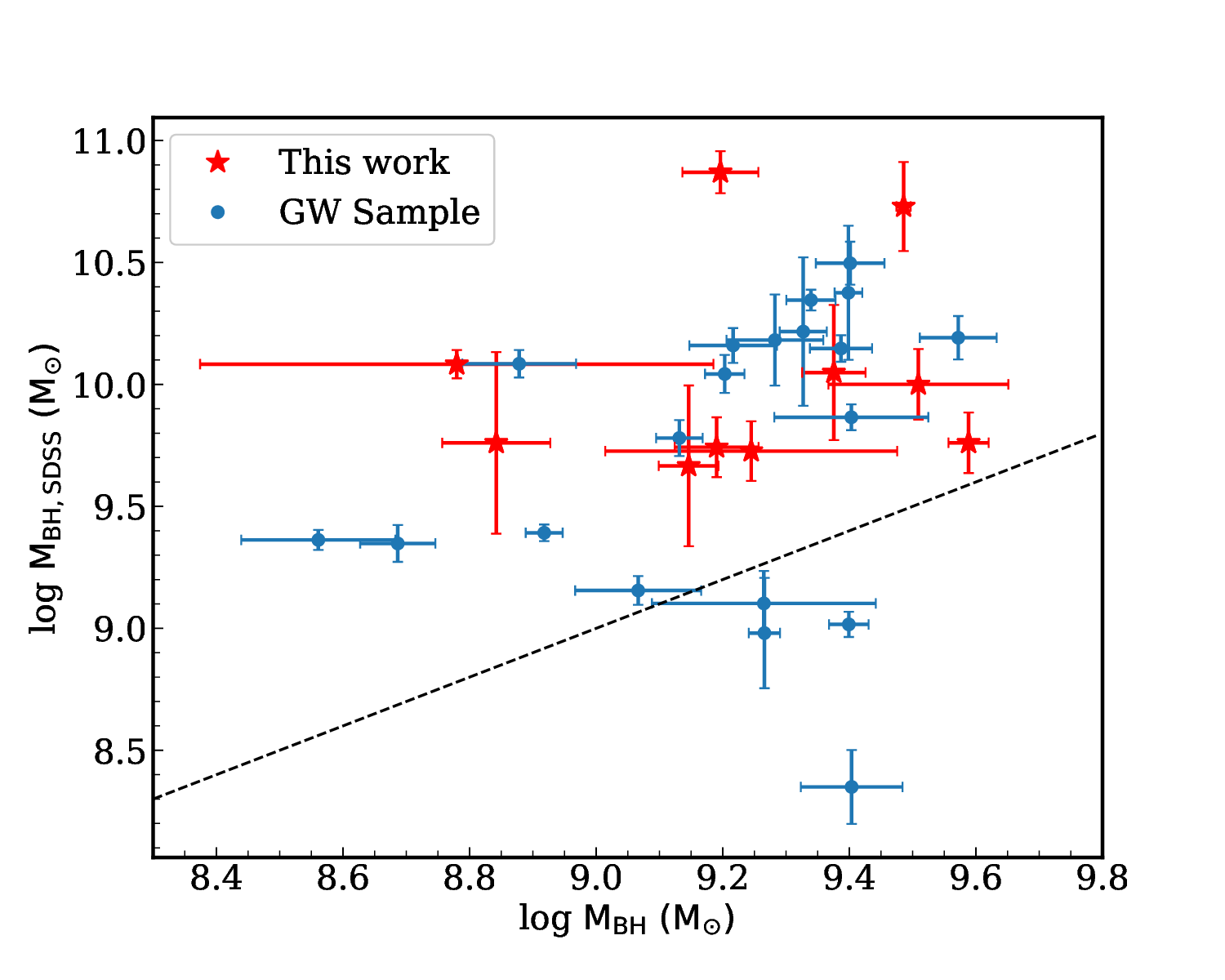}
\caption{Comparison between the H$\beta$-based virial SMBH masses derived in this study and \iona{Mg}{ii}- or \iona{C}{iv}-based virial SMBH masses in the SDSS DR7 catalog for our WLQs. 
The black dashed line is the line of equality. 
We show also the comparison for the GW sample. 
The \iona{Mg}{ii}- or \iona{C}{iv}-based virial SMBH masses are systematically larger.}
\label{fig-bhm}
\end{figure}
\begin{deluxetable}{lccc}
\tablecaption{Virial SMBH Masses and Eddington Ratios}
\tablehead{
\colhead{Source Name}&
\colhead{log~$L_{\rm 5100}$}&
\colhead{log~$M_{\rm BH}$}&
\colhead{$L_{\rm bol}/L_{\rm Edd}$} \\
\colhead{(SDSS J)}&
\colhead{($\rm erg~s^{-1}$)}&
\colhead{($\rm M_{\rm \odot}$)}&
\colhead{} \\
}
\startdata
$073306.63+462517.5$ & $46.51$ &$9.38\pm0.05$ & $0.46\pm0.06$ \\
$085344.17+354104.5$ & $46.50$ &$9.15\pm0.05$ & $0.76\pm0.08$ \\
$111401.30+222211.4$ & $46.32$ &$9.20\pm0.05$ & $0.48\pm0.07$ \\
$122709.48+310749.3$ & $46.48$ &$8.84\pm0.09$ & $1.47\pm0.29$ \\
$134601.28+585820.2$ & $46.16$ &$9.51\pm0.14$ & $0.18\pm0.06$ \\
$153412.68+503405.3$ & $46.68$ &$9.49\pm0.01$ & $0.48\pm0.01$ \\
$153714.26+271611.6$ & $46.49$ &$8.78\pm0.40$ & $1.73\pm1.62$ \\
$154329.47+335908.7$ & $46.87$ &$9.59\pm0.03$ & $0.54\pm0.04$ \\
$154503.23+015614.7$ & $46.17$ &$9.24\pm0.23$ & $0.33\pm0.17$ \\
$161245.68+511816.9$ & $46.15$ &$9.19\pm0.07$ & $0.36\pm0.05$
\enddata
\label{tab:bhml}
\tablecomments{The uncertainties are measurement uncertainties propagated from the uncertainties of the relevant parameters in Equation~\ref{eq:mbh}, and they do not include systematic uncertainties associated with the virial mass estimates.}
\end{deluxetable}

With the \hbox{rest-frame} optical spectra that cover the H$\rm \beta$ lines, we are able to estimate more reliably the \hbox{single-epoch} virial SMBH masses for our sources. We adopted the \citet{bentz2009} formula:

\begin{equation}
	\frac{M_{\rm BH}}{\rm 10^6 M_{\odot}} = 5.05 \left[ \frac{L_{5100} }{10^{44}~{\rm erg~s^{-1}}}\right]^{0.5} \left[\frac{{\rm FWHM} \left( {\rm H}\beta\right)}{10^3~{\rm km~s}^{-1}}\right]^2 ,
\label{eq:mbh}
\end{equation}

\noindent
The same formula was used in \citet{Plotkin2015} for their WLQ sample. 
The resulting SMBH masses are listed in Table~\ref{tab:bhml}. 
Our WLQs are massive SMBHs with ${\rm log}~(M_{\rm BH}/{\rm M_{\odot}})$ in the range of 8.78 to 9.59.
Nine of the ten WLQs have \iona{Mg}{ii}-based \hbox{single-epoch} virial SMBH masses in the \citet{Shen2011} 
SDSS DR7 catalog,
and the other one has a \iona{C}{iv}-based virial mass.
The comparison of the masses are displayed in Figure~\ref{fig-bhm}.
The H$\beta$-based virial SMBH masses are systematically lower,
by an average factor of 11.2 for our WLQs.
\iona{Mg}{ii}-based virial SMBH masses for typical quasars are in general consistent with H$\beta$-based ones \citep[e.g.,][]{Shen2011,Bian2012,Tilton2013,Maithil2022}.
However, most of our WLQs have weak \iona{Mg}{ii} lines,
which likely hamper accurate line-profile measurements.
Therefore, our H$\beta$-based masses should be more reliable. 

\begin{figure}
\centering \includegraphics[scale=0.35]{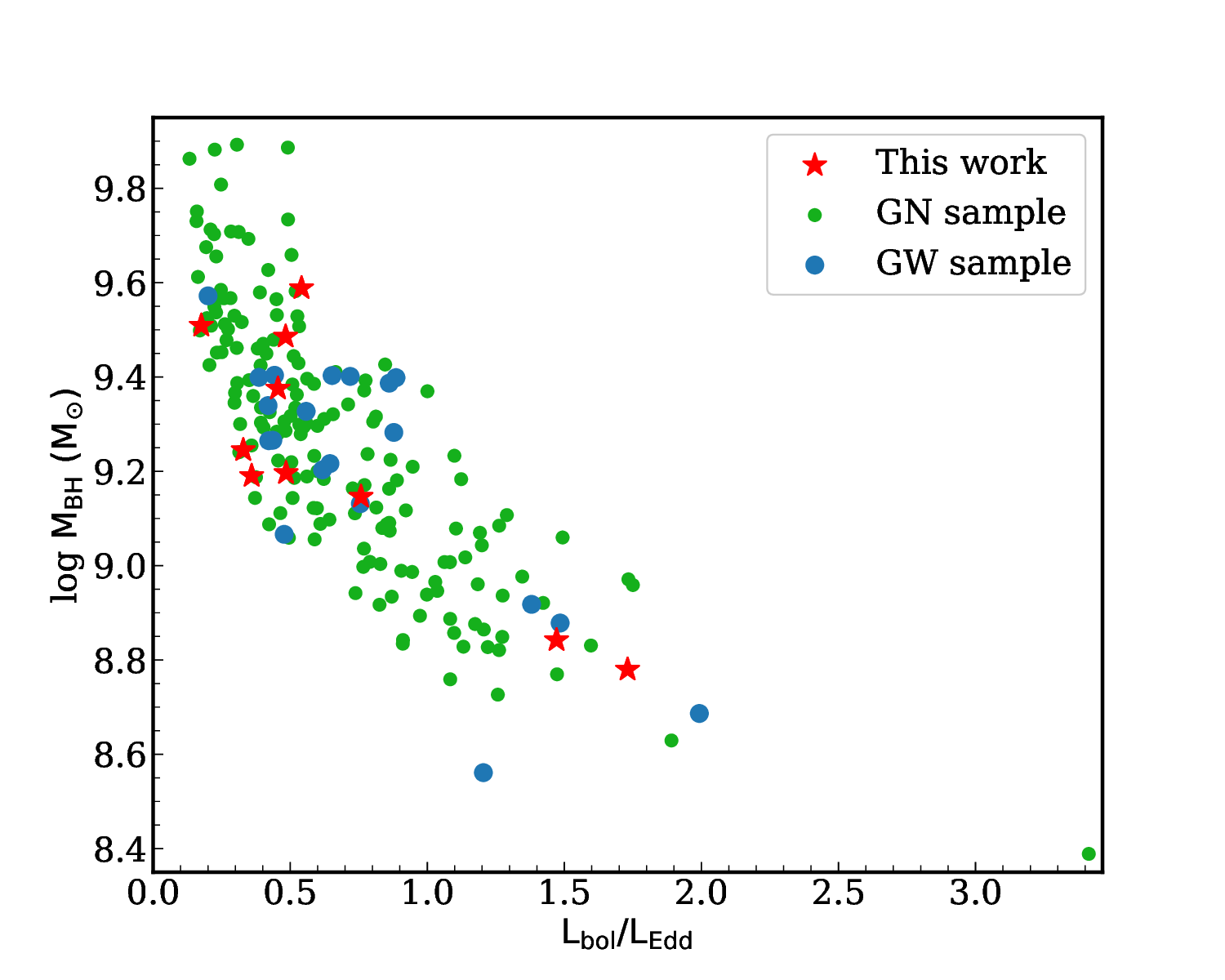}
\caption{$M_{\rm BH}$ vs. $L_{\rm bol}/L_{\rm Edd}$ for our sample and the comparison samples. 
There is no significant difference between the $M_{\rm BH}$ or $L_{\rm bol}/L_{\rm Edd}$ distributions of these samples.}
\label{fig-bhedd}
\end{figure}

Adopting the $L_{\rm 5100}$ bolometric correction prescription in \citet{Netzer2019}, we estimate the Eddington ratios with the following equation:

\begin{equation}
	L_{\rm bol}/L_{\rm Edd} = \left[f({\rm L}) \times L_{\rm 5100}\right] / \left[ 1.5\times 10^{38}~M_{\rm BH}/{\rm M_{\odot}}\right],
\label{eq:ledd}
\end{equation}

\noindent
where $f({\rm L}) = 40\times[L_{\rm 5100}/10^{42}\rm \ erg\ s^{-1}]^{-0.2}$ is a \hbox{luminosity-dependent} bolometric correction factor (Table~1 of \citealt{Netzer2019}).
The resulting Eddington ratios are listed in Table~\ref{tab:bhml}. 
Our WLQs have large Eddington ratios in general; the median $L_{\rm bol}/L_{\rm Edd}$ value is 0.5 with an interquartile range of 0.4 to 0.7. 
For comparison, we computed H$\rm \beta$-based SMBH masses and Eddington ratios for the two comparison samples using the measurements in \citet{Matthews2023}. 
Eight of the 20 GW quasars have \iona{Mg}{ii}-based masses in the SDSS DR7 catalog,
and another ten objects have \iona{C}{iv}-based masses.
For the other two quasars without DR7 masses,
we adopted their \iona{C}{iv}-based masses in the SDSS DR16 quasar catalog \citep{Wu2022}.
A comparison of the H$\rm \beta$-based SMBH masses and SDSS DR7 masses for the GW sample is included in Figure~\ref{fig-bhm}. 
The H$\rm \beta$-based virial SMBH masses are again systematically smaller.
In Figure~\ref{fig-bhedd}, we plot the $M_{\rm BH}$ versus $L_{\rm bol}/L_{\rm Edd}$ distributions for our sample and the comparison samples. 
There does not appear to be any significant difference between the $M_{\rm BH}$ or $L_{\rm bol}/L_{\rm Edd}$ distributions for our sample and the comparison samples. 
For the $M_{\rm BH}$ distributions of our sample and the GN sample, a K-S test yielded $P_{\rm null}=0.93$, and for the $L_{\rm bol}/L_{\rm Edd}$ distributions of these two samples, the $P_{\rm null}$ value is 0.74, both suggesting similar distributions. 
We also tested estimating SMBH masses for our WLQs based on their H$\alpha$ line profiles, following the approach in \citet{Greene2005}. 
The derived masses are comparable to the H$\beta$-based estimates, and the result regarding the Eddington-ratio distributions remains the same.

We caution that there are substantial systematic uncertainties associated with the $L_{\rm bol}/L_{\rm Edd}$ estimates, especially for \hbox{super-Eddington} accreting quasars. 
The virial assumption might no longer be valid for \hbox{super-Eddington} accreting quasars due to the impact of the large radiation pressure and anisotropy of the ionizing radiation \citep[e.g.,][]{Marconi2008,Marconi2009,Netzer2010,Krause2011,Pancoast2014,Li2018}. 
The radius--luminosity relation adopted in Equation~\ref{eq:mbh} might break down in the \hbox{super-Eddington} regime \citep[e.g.,][]{Hu2008,Wang20142,Du2015,Du2016}, and \citet{DU2019} have proposed a new scaling relation with ${R}_{\rm Fe\ II}$ being an additional parameter. 
We estimated $M_{\rm BH}$ based on this new scaling relation and the resulting $M_{\rm BH}$ values are smaller by a factor of $\approx~6.2$ on average. 
Subsequently, the $L_{\rm bol}/L_{\rm Edd}$ values are larger by a factor of $\approx~6.2$ on average. 
Substantial uncertainties might also come from the $L_{\rm bol}$ estimates for \hbox{super-Eddington} accreting quasars due to the uncertain extreme UV radiation \citep[e.g.,][]{Castell2016,Kubota2018}.
Moreover, the bolometric luminosity and the Eddington ratio are probably not good representatives of the accretion power in the \hbox{super-Eddington} regime, as a large fraction of the energy may be advected into the SMBH or be converted into the mechanical power of a wind \citep[e.g.,][]{Wang2014,Jiang2019}. 
Therefore, a simple comparison of the $L_{\rm bol}/L_{\rm Edd}$ distributions for the WLQ sample and the GN sample does not provide strong constraints on the accretion power of WLQs compared to typical high-$z$ quasars. 

\section{Discussion}
\label{sec:dis}

\subsection{Thick Accretion Disk and Associated Outflow Scenario for WLQs}
\label{subsec:scenario}

Our NIR spectroscopy of ten WLQs enlarges the limited samples of \iona{C}{iv}-selected WLQs that have \hbox{rest-frame} optical spectroscopy \citep[e.g.,][]{Shemmer2010,Wu2011,Plotkin2015,Shemmer2015,Ha2023}. 
Based on VLT/X-Shooter spectra of six WLQs, \citet{Plotkin2015} found that statistically the H$\rm \beta$ line emission in WLQs is not as exceptionally weak as for the \iona{C}{iv} line emission. 
Our results (Section~\ref{subsec:elp}) reinforce this finding that the \hbox{low-ionization} lines are not strongly affected in these systems. 
Comparison with the GW sample (Table~\ref{tab:Samplestats}) also indicates that this archival WLQ sample shows typical H$\beta$ emission as well. 
These results are broadly consistent with the TDO obscuration scenario, where a soft ionizing continuum is produced after the TDO shielding and only the \hbox{high-ionization} BELR is significantly affected.

\begin{figure}
\centering \includegraphics[scale=0.35]{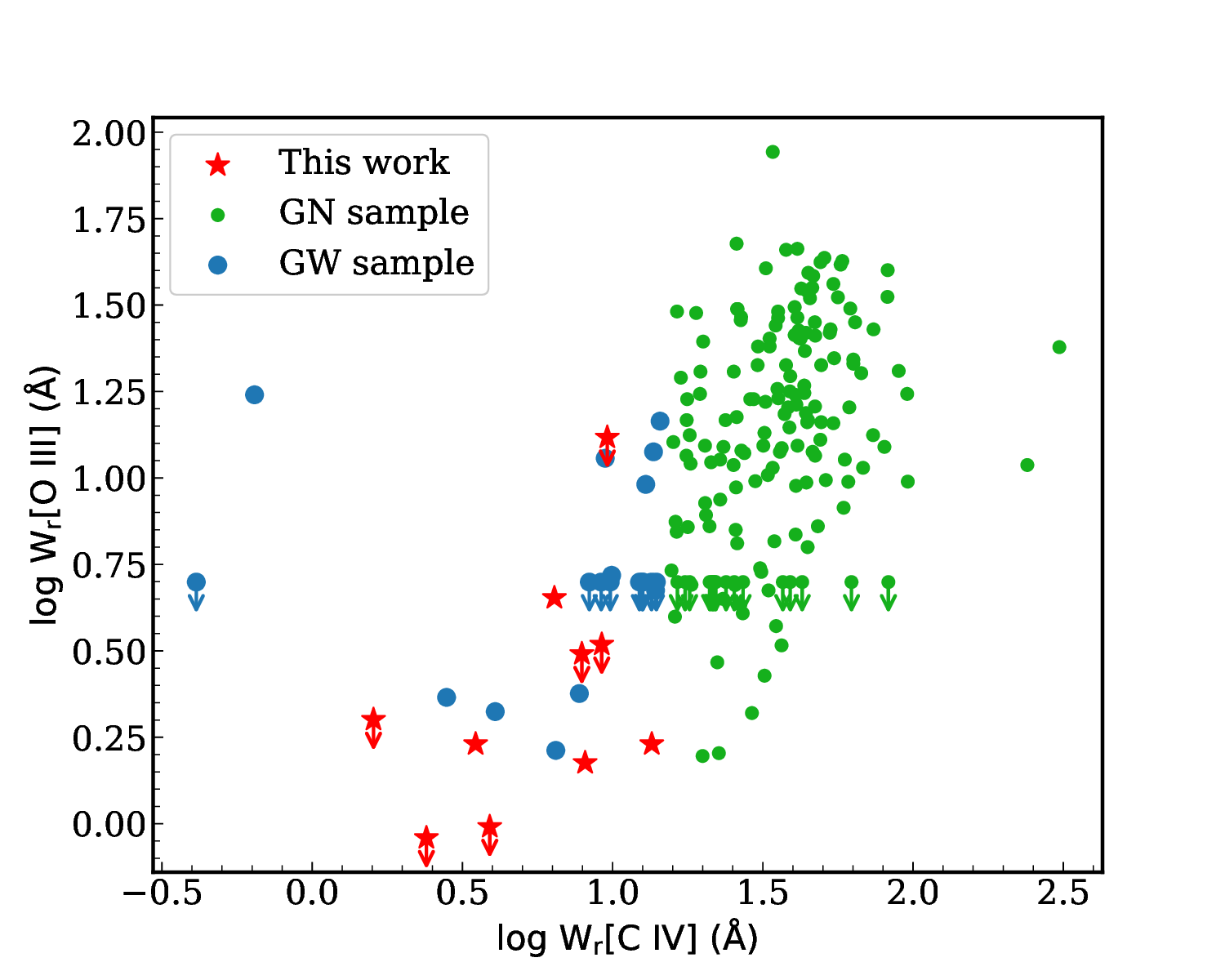}

\caption{log $W_{r}$[\iona{O}{iii}] vs. log $W_{r}$[\iona{C}{iv}] for our WLQ sample (red stars) and the comparison samples (blue and green dots).
Upper limits are indicated by the downward arrows.
The two REWs appear to be positively correlated, and the WLQ and GW quasars occupy the bottom left corner.}
\label{fig-oiiiciv}
\end{figure}

The most striking feature of the WLQ optical spectra appears to be the universally weak/absent [\iona{O}{iii}] emission, which was noticed before but did not attract much attention due to the limited samples available. 
As discussed in Section~\ref{subsec:elp}, such weak/absent [\iona{O}{iii}] emission is more likely an indicator of \hbox{super-Eddington} accretion, consistent with the TDO scenario. 
Being a narrow forbidden emission line, the [\iona{O}{iii}]~$\lambda 5007$ line is produced in the galactic-scale narrow \hbox{emission-line} region (NELR), and its strength is believed to be correlated with the ionizing ($\gtrsim35~\rm eV$) luminosity received by the NELR. 
A positive correlation between the [\iona{O}{iii}] luminosity and the hard \xray\ luminosity for active galactic nuclei (AGNs) across a broad luminosity range has been reported \citep[e.g.,][]{Heckman2005,Panessa2006,Ueda2015}. 
Since at least $\approx50\%$ of the WLQs (i.e., those showing typical levels of \xray\ emission) produce standard quasar SEDs from IR to \hbox{X-ray}, the weak [\iona{O}{iii}] emission should not be due to an intrinsically weak ionizing continuum.

There also appears to be a negative correlation, albeit with a large scatter, between the [\iona{O}{iii}] REW and quasar luminosity \citep[e.g.,][]{Sulentic2004,Stern2012,Coatman2019}; one possible explanation is that the [\iona{O}{iii}] NELR cannot grow indefinitely with the ionizing luminosity as there will be no gas available to be ionized.\footnote{However, we note that the inner boundary for the [\iona{O}{iii}] NELR might still be fairly compact even when the luminosity is high (e.g., $\sim 100$ pc for an ionizing luminosity of $\rm 10^{47}~erg~s^{-1}$), according to the computation results in Figure 6 of \citet{Stern2014}.}
However, in our case here, the WLQ and comparison samples have comparable luminosities (e.g., Figure~\ref{fig-mi_z}). 
We ran a K-S test on the bolometric luminosity distributions of the WLQ and GN samples, and the resulting $P_{\rm null}$ value is 0.15, indicating no significant difference. 
Therefore, the weaker [\iona{O}{iii}] emission of the WLQs should not be due to higher luminosities than for the GN sample.

It has also been proposed that the [\iona{O}{iii}] emission strength (and some of the Eigenvector 1 correlations) could be mainly explained by orientation effects \citep[e.g.,][]{Bisogni2017}. 
In this scenario, the [\iona{O}{iii}] emission is isotropic while the disk continuum emission is orientation dependent. 
Thus a face-on view would result in a smaller [\iona{O}{iii}] REW than that in an edge-on view, and the weak [\iona{O}{iii}] emission of the WLQs could be explained if they preferentially have small inclination angles.
However, the significant positive correlation between Eigenvector 1 and [\iona{O}{iii}] luminosity argues against an orientation dominated dependence (Figure 8 and Section 4.1 of \citealt{Boroson1992}).
Moreover, the WLQs were selected based on the weak \iona{C}{iv} emission, while in this orientation scenario, broad emission-line strength does not have an orientation dependence. 
It thus appears impossible to select a sample of face-on quasars with the \iona{C}{iv} REW criterion.
Another argument against this explanation is based on the $\approx30$ -- $50\%$ of X-ray weak WLQs observed (Section~\ref{sec:intro}).
The X-ray weakness is likely attributed to X-ray absorption, but it is hard to explain such significant X-ray absorption with face-on views.

In Figure~\ref{fig-oiiiciv}, we show the WLQ, GW, and GN samples in the log $W_{r}$[\iona{O}{iii}] vs. log $W_{r}$[\iona{C}{iv}] plane.
It is clear that the WLQ and GW quasars are located in the bottom left corner, with both weak [\iona{O}{iii}] and \iona{C}{iv} emission.
The two REWs appear to be positively correlated, with a Spearman rank-order correlation coefficient of $\rho=0.556$.
Similar to the [\iona{O}{iii}] emission, the strength of the \iona{C}{iv} emission also reflects the ionizing ($\gtrsim 48$~eV) luminosity received by the \iona{C}{iv} BELR.
In the TDO scenario for WLQs, the equatorial \iona{C}{iv} BELR is shielded by the TDO, leading to the weak line emission.
A similar mechanism might work for the weak [\iona{O}{iii}] emission in WLQs.
The ionizing extreme UV and X-ray radiation is confined to the funnel region in the configuration of a geometrically thick accretion disk (see Figure 1 of \citealt{Ni2018}), and thus the [\iona{O}{iii}] NELR has a small covering factor (i.e., a small solid angle for the ionization cone; e.g., \citealt{Wang2014}), resulting in the weak line emission.
Larger samples of WLQs with [\iona{O}{iii}] measurements will help to constrain statistically the covering factors and photoionization conditions of the NELRs \citep[e.g.,][]{Baskin2005}.

Strong optical \iona{Fe}{ii} emission appears to be another important feature, although the contrast between WLQs and typical quasars is not as high as that for the [\iona{O}{iii}] emission (Table~\ref{tab:Samplestats}). 
This feature also suggests high accretion rates in general, and thus it is consistent with the TDO scenario. 
Although many studies have suggested a positive correlation between \iona{Fe}{ii} emission strength and Eddington ratio, there is no clear understanding of the underlying physics, mainly due to the complex excitation mechanisms and the poorly constrained location for the \iona{Fe}{ii} emission \citep[e.g.,][]{Dong2011,Marinello2016}. 
One proposed interpretation is that only high-density clouds are gravitationally bound in \hbox{super-Eddington} accreting systems, which produce strong Fe emission due to radiative transfer effects \citep{Sameshima2011}. 
Another possibility is that \hbox{super-Eddington} accreting AGNs might have high metal contents \citep[e.g.,][]{Sniegowska2021,Denimara2024,Marziani2024}.

Overall, the optical spectral properties of our systematically selected sample of WLQs provide further support to the TDO scenario. 
With the H$\beta$ \hbox{emission-line} measurements, we are able to estimate more reliably the SMBH masses. 
However, as discussed in Section~\ref{subsec:bh}, extra caution is needed when assessing the accretion power for \hbox{super-Eddington} accretion with the Eddington-ratio parameter. 

WLQs also display exceptional \xray\ properties. 
Four of our sources have previous \xray\ coverage from \citet{Wu2012}, \citet{Luo2015}, or \citet{Ni2022}.
Two of them show typical levels of \xray\ emission (\xray\ normal) as expected from the $\alpha_{\rm OX}\textrm{--}L_{2500~{\textup{\AA}}}$ relation, while the other two were not detected in the Chandra observations and are \xray\ weak by factors of $>10.1$ and $>27.4$ compared to the $\alpha_{\rm OX}\textrm{--}L_{2500~{\textup{\AA}}}$ relation. 
The basic \xray\ and optical \hbox{emission-line} properties for these four WLQs are listed in Table~\ref{tab:x_ray}. 
Under the TDO scenario, \xray\ normal and \xray\ weak WLQs are intrinsically similar, but the \xray\ weak ones are viewed at larger inclination angles with TDO obscuration along the line of sight. 
Our sample of four objects is too small for statistical comparisons, but among the \hbox{emission-line} properties studies here, the \xray\ weak WLQs appear to have broader H$\beta$ lines, consistent with the expectation from larger inclination angles given a flattened BELR geometry \citep[e.g.,][]{Wills1986,Runnoe2013,Shen2014}. 
We caution that the H$\beta$ width should have considerable object-to-object scatter, and statistical assessment on a larger sample is needed to confirm such a difference.

Another interesting question to explore is the connection between WLQs and \hbox{super-Eddington} accreting quasars in general. 
For example, are WLQs special among \hbox{super-Eddington} accreting quasars? 
There are small samples of \hbox{super-Eddington} accreting quasars selected based on their Eddington ratios, luminosities, \xray\ properties (e.g., steep spectral shapes), and/or Eigenvector 1 parameters \citep[e.g.,][]{Nardini2019,Liu2021, Lusso2021,Laurenti2022}. 
These objects share many similarities with WLQs in terms of their SEDs (typical IR--UV quasar SEDs), \xray\ properties (sometimes \xray\ weak and \xray\ variable), and optical spectral properties (as discussed in this study). 
Some of these quasars indeed show weak \iona{C}{iv} emission lines and could be considered WLQs, but some show nominal levels of \iona{C}{iv} emission. 
Thus \hbox{super-Eddington} accreting quasars are unlikely all WLQs. 
Under the TDO scenario, the weakness of \iona{C}{iv} line is controlled by the TDO shielding. 
The covering factor and/or column density of the TDO in a strong \iona{C}{iv} emitter are thus lower than those of a WLQ; WLQs are probably extremely \hbox{super-Eddington} accreting with high-covering-factor high-density TDOs. 
Observationally, one could test this hypothesis via comparing the fraction of \xray\ weak quasars and the \xray\ weakness factors between these two populations. 
Examining these two populations in the Eigenvector~1 parameter space might also provide insights. 

\begin{deluxetable*}{lccccccc}
\tablecaption{\xray\ and Optical Emission-Line Properties}
\tablehead{
\colhead{Source Name}                   &
\colhead{$\alpha_{\rm OX}$}                   &
\colhead{$\Delta\alpha_{\rm OX}$}     &
\colhead{$f_{\rm weak}$} &
\colhead{$\rm H \beta$~REW}&
\colhead{[\iona{O}{iii}]~REW}&
\colhead{\iona{Fe}{ii}~REW}&
\colhead{$\rm H \beta$~FWHM}              \\
\colhead{(SDSS J)}   &
\colhead{}   &
\colhead{}   &
\colhead{}   &
\colhead{($\rm \textup{\AA}$)}&
\colhead{($\rm \textup{\AA}$)}&
\colhead{($\rm \textup{\AA}$)}&
\colhead{($\rm km~s^{-1}$)} \\
\noalign{\smallskip}\hline\noalign{\smallskip}
\multicolumn{8}{c}{\xray\ Weak Quasars}
}
\startdata
$       153412.68+503405.3$&$ <-2.11$&$<-0.39$&$ >10.1$&$91.1\pm1.5$&$<3.3$& $59\pm3$&$5040\pm65$\\
$       134601.28+585820.2$&$ <-2.22$&$<-0.55$&$  >27.4$&$17.3\pm1.3$&$<0.91$& $56\pm5$&$6990\pm1100$\\
\noalign{\smallskip}\hline\noalign{\smallskip}
\multicolumn{8}{c}{\xray\ Normal Quasars} \\
\noalign{\smallskip}\hline\noalign{\smallskip}
$       153714.26+271611.6$&$ -1.82$ &$-0.06$&$  1.4  $&$24.9\pm0.4$&$<3.1$& $50\pm2$&$2500\pm1200$\\
$       161245.68+511816.9$&$ -1.67$ &$0.02$  &$ 0.9  $&$25.4\pm1.4$&$<2.0$& $70\pm4$&$4870\pm360$
\enddata
\tablecomments{The \xray\ properties are adopted from \citet{Wu2012}, \citet{Luo2015}, or \citet{Ni2022}. $\Delta\alpha_{\rm OX}$ is defined as the difference between the observed $\alpha_{\rm OX}$ value and the one expected from  
the $\alpha_{\rm OX}\textrm{--}L_{2500~{\textup{\AA}}}$ relation ($\Delta\alpha_{\rm OX}=\alpha_{\rm OX}-\alpha_{\rm OX,exp}$), and $f_{\rm weak}=10^{-\Delta\alpha_{\rm OX}/0.383}\approx 403^{-\Delta\alpha_{\rm OX}}$. 
The \hbox{emission-line} properties are the same as those in Table~\ref{tab:elp}.}
\label{tab:x_ray}
\end{deluxetable*}

\subsection{Candidate Extreme \texorpdfstring{[\iona{O}{iii}]~$\lambda5007$}{oiii} Outflows in \texorpdfstring{SDSS~J$153714.26+271611.6$}{name}}
\label{subsec:outflow}

The spectrum of SDSS~J$153714.26+271611.6$ (J1537 hereafter) suggests the presence of blueshifted broad [\iona{O}{iii}] emission lines (Figure~\ref{fig-spec1}). 
Two broad components are required to fit the spectrum, with FWHMs of $\approx 1910\pm140$ and $\rm \approx~2900\pm260~km~s^{-1}$.
If these two components are the [\iona{O}{iii}]~$\lambda4959$ and $\lambda5007$ doublet, the blueshifts are $\approx~500$--$2000~\rm km~s^{-1}$.
However, the line fluxes do not follow the $1:3$ ratio, suggesting that these might both be the [\iona{O}{iii}]~$\lambda5007$ line but at different blueshifts. 
The blueshifts are thus $\approx~500\pm170$ and $\approx~4900\pm140$ $\rm km~s^{-1}$.

There is no previous report of blueshifted broad [\iona{O}{iii}] emission lines in WLQs, and there are only a limited number of such cases in quasars \citep[e.g.,][]{Brusa2015,Brusa2016,Zakamska2016,Bischetti2017,Xu2020,Fukuchi2023}. 
These lines suggest strong [\iona{O}{iii}] outflows in the host galaxy, and they provide strong evidence for quasar feedback. 
Super-Eddington accreting quasars are expected to produce powerful \hbox{accretion-disk} winds \citep[e.g.,][]{Giustini2019,Jiang2019}, which may develop into massive galactic-scale outflows. 
The [\iona{O}{iii}] blueshifts in J1537 indeed appear higher than those discovered in other quasars \citep[e.g.,][]{Zakamska2016}.  
Under the TDO scenario, the [\iona{O}{iii}] ionizing cone has a small solid angle. 
Thus blueshifted lines indicate a small inclination angle. 
J1537 shows a nominal level of \xray\ emission (Table~\ref{tab:x_ray}), which also indicates a relatively small inclination angle.
Such an \hbox{inclination-angle} constraint might also limit the occurrence of blueshifted broad [\iona{O}{iii}] lines in WLQs in general.
Nevertheless, we caution that the P200/TSpec spectrum has a limited SNR, and there might also be contamination from additional \iona{Fe}{ii} emission around [\iona{O}{iii}] \citep[e.g.,][]{Kova2010,Bischetti2017}.
In certain instances, these lines can be incorrectly identified as [\iona{O}{iii}] emission, even though they don't align well with the characteristics of these two broad components.
Therefore, we consider J1537 a candidate for containing extreme [\iona{O}{iii}]~$\lambda5007$ outflows.
Higher quality NIR spectra are required to confirm the presence of these extreme blueshifted broad [\iona{O}{iii}] lines and study their kinematics.

\section{Summary}
\label{sec:sum}
In this paper, we present systematic investigations of the optical \hbox{emission-line} properties for a sample of $z\sim2$ WLQs observed with P200/TSpec. 
The final sample contains ten WLQs with \iona{C}{iv} REWs $< 15~\textup{\AA}$ (nine having \iona{C}{iv} REWs $< 10~\textup{\AA}$).
Four sources have previous \xray\ coverage.
For comparison, we constructed a typical quasar (GN) sample containing 168 high-redshift ($\rm z\approx~1.5~\textrm{--}~3.5$) quasars with \hbox{GNIRS-DQS} NIR spectra \citep{Matthews2021}. 
The 20 archival WLQs in the \citet{Matthews2023} \hbox{GNIRS-DQS} catalog constitute the GW comparison sample; the \hbox{emission-line} properties of our sample and the GW sample are consistent overall. 

The \hbox{emission-line} features we examined include the H$\beta$ REW, H$\beta$ FWHM, [\iona{O}{iii}] REW, and ${R}_{\rm Fe\ II}$ (Figure~\ref{fig-elp}).
We confirmed the previous finding that the WLQ H$\beta$ lines, as a major \hbox{low-ionization} line, are not significantly weak compared to typical quasars. 
This result supports the soft ionizing continuum scenario where only the \hbox{high-ionization} BELR is strongly affected, leading to the exceptionally weak \iona{C}{IV} emission. 
There is no significant difference between the H$\beta$ FWHM distributions either. 
The most prominent feature of the WLQ optical spectra is the universally weak/absent [\iona{O}{iii}] emission. 
WLQs also display stronger optical \iona{Fe}{ii} emission (${R}_{\rm Fe\ II}$) than typical quasars. 
Weak [\iona{O}{iii}] emission and strong \iona{Fe}{ii} emission suggest high accretion rates for the WLQs considering the Eigenvector~1 correlations. 
See Section~\ref{subsec:elp}.

We derived H$\beta$-based \hbox{single-epoch} virial SMBH masses and Eddington ratios for our objects. 
These are massive SMBHs ($\sim10^9~\rm M_{\odot}$) with large Eddington ratios (median $L_{\rm bol}/L_{\rm Edd}=0.5$). 
There is no significant difference between the $M_{\rm BH}$ or $L_{\rm bol}/L_{\rm Edd}$ distributions for our sample and the comparison samples, 
but extra caution is needed for the interpretation. 
See Section~\ref{subsec:bh}.

Our results provide further support to the TDO scenario, where WLQs are \hbox{super-Eddington} accreting and the weak \hbox{high-ionization} UV emission lines are produced due to a soft ionizing continuum from TDO shielding. 
This scenario can also explain the remarkable \xray\ properties and other multiwavelength properties of WLQs. 
The candidate extreme [\iona{O}{iii}] $\lambda5007$ outflows (blueshifts of $\approx~500$ and $\rm \approx~ 4900~km~s^{-1}$) in J1537 might connect to the small-scale TDO. 
See Section~\ref{sec:dis}. 

WLQs appear to be a unique quasar population that is able to provide important clues about SMBH accretion, \hbox{accretion-driven} outflows, and the nuclear gaseous environment. 
The current samples of \iona{C}{IV}-selected WLQs with \hbox{rest-frame} optical spectra are still small, and they lack systematic selections or spectral analyses.
NIR spectroscopy of a large sample uniformly selected from the SDSS quasar catalog (e.g., DR16; \citealt{Wu2022}), aided with \xray\ coverage,
will be helpful for constraining more reliably their nature.
We note that although \citet{Ni2022} has composed a well-defined sample of 32 WLQs with good \xray\ coverage,
the redshifts for a large fraction of these objects fall outside the observing window for ground observations due to telluric absorption.
NIR spectroscopy from JWST would be very helpful in this regard.

~\\

We thank J.~M.~Wang for helpful discussions.
Y.C. and B.L. acknowledge financial support from the National Natural Science Foundation of China grants 11991053 and 11991050, China Manned Space Project grants NO. \hbox{CMS-CSST-2021-A05} and NO. \hbox{CMS-CSST-2021-A06}, and CNSA program D050102.
WNB acknowledges the Eberly Endowment at Penn State.

This research uses data obtained through the Telescope Access Program (TAP), which has been funded by the Strategic Priority Research Program “The Emergence of Cosmological Structures” (Grant No. XBD09000000), National Astronomical Observatories, Chinese Academy of Sciences, and the Special Fund for Astronomy from the Ministry of Finance. 
Observations obtained with the Hale Telescope at Palomar Observatory were obtained as part of an agreement between
the National Astronomical Observatories, Chinese Academy of
Sciences, and the California Institute of Technology.

\bibliographystyle{aasjournal}
\bibliography{ms}

\clearpage
\appendix

\section{SDSS and P200 Spectra of \texorpdfstring{SDSS~J$164302.03+441422.1$}{name}}
\label{sec:appendix}
\begin{figure*}
\figurenum{A1}
\centering \includegraphics[scale=0.51]{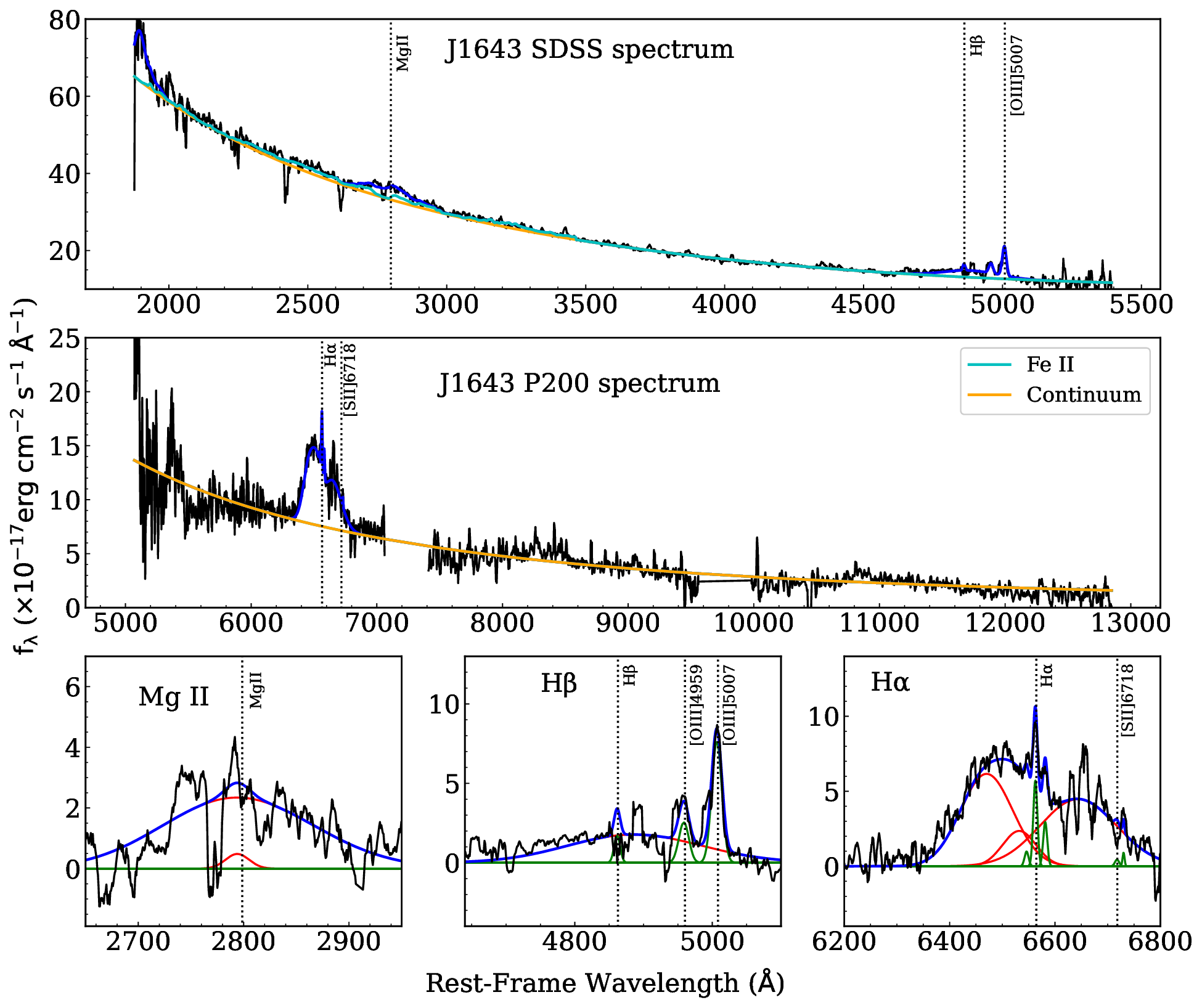}
\caption{Similar to Figure~\ref{fig-spec1} but for the SDSS and P200/TSpec spectra of SDSS~J$164302.03+441422.1$. 
In the bottom panel, we show the zoomed-in view of the spectrum in the \iona{Mg}{ii}, H$\beta$, and H$\alpha$ regions, respectively.
}
\label{fig-1643}
\end{figure*}

The \iona{Mg}{ii} line in the SDSS spectrum of SDSS~J$164302.03+441422.1$ was wrongly identified as the \iona{C}{IV} line in the SDSS~DR7 quasar catalog \citep{Shen2011}, and thus the redshift of 1.650 was incorrect. 
The wrong redshift was fixed (updated $z=0.917$) in the updated SDSS quasar catalog of \citet{Wu2022}. 
The P200/TSpec spectrum thus covers only the H$\alpha$ line. 
We fitted the SDSS and P200 spectra with the updated redshift, and the results are shown in Figure~\ref{fig-1643}. 
The REWs of the \iona{Mg}{ii}, H$\beta$, and [\iona{O}{iii}] lines are $\rm 12.56\pm1.6~{\textup{\AA}}$, $\rm 30.4\pm4.6~{\textup{\AA}}$, and $\rm 11.56\pm1.6~{\textup{\AA}}$, respectively, which are consistent with the measurements in \citet{Wu2022}. 
The H$\beta$ line appears very broad (FWHM $\approx~13500 \rm~km~s^{-1}$). 
Both the \iona{Mg}{ii} and H$\beta$ lines are relatively weak; the \iona{Mg}{ii} REW is similar to the average value for the \citet{Plotkin2015} WLQ sample, and the H$\beta$ REW is slightly smaller than the average value for our sample. 
However, the [\iona{O}{iii}] line has a typical REW compared to typical quasars (e.g., compared to the distribution in Figure~\ref{fig-elp}a). 
Since this quasar is at a lower redshift, and it does not have measurements of the \iona{C}{iv} line, we consider it only a WLQ candidate and did not include it in our sample.

\end{document}